\documentstyle [preprint,revtex,eqsecnum,epsf]{aps}

\global\arraycolsep=2pt			

\makeatletter
\def\footnotesize{\@setsize\footnotesize{9.5pt}\xpt\@xpt
\abovedisplayskip 10pt plus2pt minus 5pt
\belowdisplayskip \abovedisplayskip
\abovedisplayshortskip \z@ plus 3pt
\belowdisplayshortskip 6pt plus 2pt minus 2pt
\def\@listi{\topsep 6pt plus 2pt minus 2pt
\parsep 3pt plus 2pt minus 1pt \itemsep \parsep}}

\skip\footins 16pt plus 4pt minus 4pt
\def\fnum@figure{Figure \thefigure.}
\def\figure{\@float{figure}}
\let\endfigure\end@float
\@namedef{figure*}{\@dblfloat{figure}}
\@namedef{endfigure*}{\end@dblfloat}
\makeatother

\begin {document}

\preprint {UW/PT-92-07}

\begin {title}
    {%
    Large-order perturbation theory for the electromagnetic \\
    current-current correlation function
    }%
\end {title}

\author {Lowell S. Brown, Laurence G. Yaffe, and Chengxing Zhai}

\begin {instit}
    {%
    Department of Physics, FM-15,
    University of Washington,
    Seattle, Washington 98195
    }%
\end {instit}

\begin {abstract}
    {%
    The constraints imposed by asymptotic freedom and analyticity on the
    large-order behavior of perturbation theory for the electromagnetic
    current-current correlation function are examined.
    By suitably applying the renormalization group,
    the coefficients of the asymptotic expansion in the deep Euclidean
    region may be expressed explicitly in terms of the
    perturbative coefficients of the Minkowski space discontinuity
    (the $R$-ratio in $e^+ e^-$ scattering).
    This relation yields a ``generic'' prediction for the large-order
    behavior of the Euclidean perturbation series and suggests the
    presence of non-perturbative $1/q^2$ correction in the Euclidian
    correlation function.
    No such ``generic'' prediction can  be made for the physically
    measurable $R$-ratio.
    A novel functional method is developed to obtain these results.

    \ifpreprintsty
    \vbox to \vsize
	{%
	\vfill
	\baselineskip .28cm
	\par
	\font\eightrm = cmr8
	\eightrm \noindent
	This report was prepared as an account of work sponsored by the
	United States Government.
	Neither the United States nor the United States Department of Energy,
	nor any of their employees, nor any of their contractors,
	subcontractors, or their employees, makes any warranty,
	express or implied, or assumes any legal liability or
	responsibility for the product or process disclosed,
	or represents that its use would not infringe privately-owned rights.
	By acceptance of this article, the publisher and/or recipient
	acknowledges the U.S. Government's right to retain a non-exclusive,
	royalty-free license in and to any copyright covering this paper.%
	}%
    \pagestyle {empty}
    \newpage
    \fi
    }%
\end {abstract}

\narrowtext

\section {Introduction}

    The renormalization group relates changes in the value of the
renormalization point to equivalent changes in the values
of renormalized couplings.
For any dimensionless physical quantity depending on a single
momentum variable,
${\cal F}(q)$,
the renormalization group may be used to reexpress
a perturbative expansion with a fixed coupling $g^2(\mu^2)$
and momentum dependent coefficients,
\begin {equation}
    {\cal F} (q) \sim \sum_n \> c_n (q^2/\mu^2) \, g(\mu^2)^{2n} \,,
\end {equation}
as an asymptotic expansion%
\footnote
    {%
    We use the symbol
    $\sim$ to denote an asymptotic series in the precise sense employed by
    mathematicians: $f(z) \sim \sum\nolimits_n f_n z^n $ if, for any $N$,
    $f(z) - \sum\nolimits_{n=0}^N f_n z^n = {\cal O}(z^{N+1})$ as $z\to 0$.
    }
in powers of a running coupling $g^2(q^2)$
with fixed (purely numerical) coefficients,
\begin {equation}
    {\cal F} (q) \sim \sum_n \> c_n (1) \, g(q^2)^{2n} \,.
\label {pert-exp}
\end {equation}
In this form, one is using the renormalization group
to relate changes in the magnitude of $q^2$ to changes in the
value of the running coupling.

    The renormalization group can also be used to express a change in
the phase of $q^2$ as an equivalent change in the complex value of the
running coupling.
Consequently, when combined with analyticity, the renormalization group
may be used to relate a Euclidean space perturbation series
(in powers of $g^2(q^2)$, with $q^2$ real and positive)
to the corresponding Minkowski space perturbation series
(in powers of $g^2(-q^2)$).

    In this paper, using only the renormalization group and analyticity,
we examine the precise relation between the Euclidean space asymptotic
expansion of the electromagnetic current-current correlation function
(the ``dispersive part'')
and the corresponding asymptotic behavior of its Minkowski space
discontinuity, which is the $R$-ratio in $e^+e^-$ annihilation (the
``absorptive part'').
We find that the exact relation between these expansions
takes an extremely simple form when expressed in terms of a
natural generalization of the Borel transform of the perturbative series.
The implications of this relation on the possible large order behavior
of perturbation theory are discussed.
In particular, we find that the perturbative coefficients of the
Euclidean correlation function and the Minkowski discontinuity
need not show the same behavior at large orders.
The renormalization group and analyticity constraints alone
are {\em not\/} sufficient to uniquely determine the large order
behavior of perturbation theory, in contradiction to
a recent claim by G. West \cite {West}.
Moreover, as discussed in detail in Appendix A, the asymptotic forms
presented by West are inconsistent with the renormalization group
and do not obey the exact relation between the dispersive and
absorptive parts which we derive.

    Our general renormalization group relation implies that the
Borel transform of the Euclidean perturbation series will
exhibit an infinite set of regularly-spaced singularities.
These include the infrared and ultraviolet ``renormalon'' singularities
found from the examination of individual Feynman diagrams
\cite {tHooft,Zinn-Justin,Parisi,Mueller}.
However, for QCD the renormalization group analysis also suggests the presence
of one further singularity
which does not correspond to an expected renormalon singularity.
As we shall discuss, this extra singularity may be interpreted as indicating
the presence of non-perturbative $1/q^2$ corrections in the Euclidean
correlation function.
Such corrections will have significant implications (which we do not explore)
for phenomenological applications of the operator product expansion
such as QCD sum-rules and heavy quark expansions.
This new Borel transform singularity must be present {\em unless\/}
the perturbative coefficients of the $R$-ratio conspire to produce
an exact cancellation in a (modified) Borel transform of the
absorptive series.
Whether or not this cancellation might occur is not known;
however, as we discuss in  Appendix D, no hint of this cancellation
is seen using the known terms in the expansion of the $R$-ratio.

    In an earlier paper \cite {Brown&Yaffe}, the special case of a theory
with a one-term beta function was examined.
In this paper, we show how to extend the analysis of \cite {Brown&Yaffe}
to the case of a general beta function,
and we examine the applications of these results at greater length.
A detailed summary and discussion of our results appears in the next section.
This starts with a review of the perturbative expansion of the electromagnetic
current-current correlation function, a brief summary of the
one-term beta function results in \cite {Brown&Yaffe}, and a discussion
of the choice of renormalization scheme which will be
most convenient for the general case.
We introduce a ``modified'' Borel transform and summarize
its properties and its connection with the ordinary Borel transform.
Our results on the relation between Euclidean and Minkowski space
asymptotic behavior are then presented, followed by a discussion
of the implications of this relation on the possible large order behavior
of the perturbative series.
The connection between our work and previous renormalon results is
described, followed by a review of the relation between Borel
transform singularities and the operator product expansion.
Finally, we discuss the significance of our ``extra'' renormalon singularity
and its implications concerning non-perturbative $1/q^2$ corrections
in the Euclidean correlation function.

    The actual derivation of our results follows this lengthy summary.
By using a functional approach,
we find a very simple form for the exact solution to the
renormalization group equations relating the perturbative coefficients.
This (slightly abstract) approach exploits familiar quantum-mechanical
techniques. In this setting,
the modified Borel transformation emerges naturally as the representation
of the series in a convenient overcomplete basis, just as the usual Borel
transform corresponds to the standard coherent state representation of
the series. Some details concerning these transforms are relegated to
Appendix B.
Alternative methods (more traditional but less convenient)
for deriving the same results are sketched in Appendix C.
The calculations of the perturbative coefficients of the
current-correlation function are scattered throughout the literature.
These results are collected and summarized in our notation in Appendix
\ref {known} to enable any comparison with the asymptotic results of this paper
that the reader may wish to make.

\section {Summary of Results and Discussion}
\label {Summary}

    The electromagnetic current-current correlation function
\begin {eqnarray}
    K^{\mu\nu} (q)
    &\equiv&
	i \int (d^4x) \> e^{-iqx} \,
	\langle 0| {\cal T}  ( j^\mu(x) j^\nu(0) ) |0\rangle
\nonumber
\\
    &=&
	(g^{\mu\nu} q^2 - q^\mu q^\nu) \, K (-q^2)
\label {K_mu_nu}
\end {eqnarray}
involves a single scalar function $K(t)$
which is analytic in the entire $t = -q^2$ plane save
for a cut along the positive real axis.
The discontinuity across this cut is related to the high-energy limit
of the total $e^+e^-$ hadronic cross section if one neglects the $Z^0$
exchange contribution.
In terms of the $R$-ratio, defined as the ratio of the total cross section
for $e^+e^- \to \hbox{hadrons}$ to that for $e^+e^- \to \hbox{muon pairs}$,
we have
\begin {equation}
    R(s) = 12 \pi \, {\rm Im} \, K (s + i 0^+) \,.
\end {equation}

    In an asymptotically free theory such as QCD,
renormalized perturbation theory, plus the renormalization group,
may be used to compute the asymptotic behavior of $K(t)$ as $|t| \to \infty$.
The discontinuity ${\rm Im} \, K(s+i0^+)$ has an asymptotic
expansion\footnote{We assume that $K(t)$ does not contain unexpected,
pathological terms such as $\exp\{ i \sqrt{t/\Lambda^2} \}$ which,
away from the positive real axis, make no contribution to the
asymptotic expansion (\ref{K-exp}).} in
powers of the running coupling $g^2(s)$,
\begin {equation}
    {\rm Im} \, K(s+i0^+) \sim
    \sum_{n=0}^\infty \> a_n \, g^{2n}(s) \,,
\label {Im K-exp}
\end {equation}
while the asymptotic behavior along the negative real $t$-axis
(corresponding to the Euclidean space correlation function) is given
by
\begin {equation}
    K (t) \sim \kappa(\mu^2)
    + \tilde c_{-1} \, g^2(-t)^{-1}
    + \tilde c_0 \, \ln g^2(-t)
    + \sum_{n=1}^\infty \> c_n \, g^{2n}(-t) \,.
\label {K-exp}
\end {equation}
 The origin of the non-analytic $1/g^2$ and $\ln g^2$ terms is
reviewed in Sec.~\ref {ren group}.  Only the constant term
$\kappa(\mu^2)$ depends on the renormalization point $\mu$; the
perturbative coefficients $\{ \tilde c_{-1}, \tilde c_0, c_n \}$ and
$\{ a_n \}$ are pure numbers, independent of $\mu^2$ and all mass
parameters.  All mass-dependent terms, such as $m^2 /t$, vanish faster
as $|t| \to \infty$ than any power of the running coupling, and hence
they may be neglected.  We will refer to the coefficients $\{ a_n \}$
in the expansion of the discontinuity as the ``absorptive''
coefficients, and the coefficients $\{ \tilde c_{-1}, \tilde c_0, c_n
\}$ in the Euclidean expansion as the ``dispersive'' coefficients.

    The absorptive and dispersive coefficients are not independent;
one may express the absorptive coefficients in terms of dispersive
coefficients, or vice-versa.  The precise relation between the
coefficients depends on the behavior of the running coupling as
determined by the beta function
\begin {equation}
    \beta (g^2(\mu^2)) \equiv \mu^2 \, {d g^2 (\mu^2) \over d \mu^2}
\,,\qquad
\label {beta-function}
\end {equation}
with the explicit form governed by the coefficients of its
perturbative expansion,
\begin {equation}
    \beta (g^2) \sim -b_0 \> g^4 -b_1 \> g^6 -b_2 \> g^8 - \cdots \,.
\label {beta-function-expansion}
\end {equation}

\subsection {One-term beta function results}

    In the earlier paper \cite {Brown&Yaffe}, the special case of a
one-term beta function,
\begin {equation}
    \beta (g^2) = -b_0 \, g^4 \,,
\end {equation}
 was studied.  The results are remarkably simple when expressed in
terms of the Borel transforms of the perturbative coefficients,
defined as
\begin {eqnarray}
    A(z) &\equiv& \sum_{n=1}^\infty \> {n \, a_n \over \Gamma(n{+}1)}
\> z^n
\label {1-term A(z)} \,,
\\
\noalign {\hbox {and}}
    C(z) &\equiv& \tilde c_0 +
    \sum_{n=1}^\infty \> {n \, c_n \over
\Gamma(n{+}1)} \> z^n \,.
\label {1-term C(z)}
\end {eqnarray}
 As shown in \cite {Brown&Yaffe}, these satisfy
\begin {equation}
    A(z) = \sin (\pi b_0 z) \, C(z) \,.
\label {A=sin C}
\end {equation}
Expanding both sides of this result in powers of $z$ generates
explicit expressions for the absorptive coefficients as linear
combinations of the dispersive coefficients,
\begin {eqnarray}
    a_{2n} &=& \sum_{k=0}^{n-1} \>
    {(-)^k \> (2n{-}1)!
\over (2k{+}1)! \> (2n{-}2k{-}2)!} \;
(\pi b_0)^{2k+1} \,
c_{2(n-k)-1} \,,
\label {a_even} \\
\noalign {\hbox{and}}
    a_{2n+1} &=& \sum_{k=0}^{n-1} \>
    { (-)^k \> (2n)! \over
(2k{+}1)! \> (2n{-}2k{-}1)!} \;
(\pi b_0)^{2k+1} \,
c_{2(n-k)}
\> + \> {(-)^n \over 2n{+}1} \>
(\pi b_0)^{2n+1} \, \tilde c_0 \,.\qquad
\label {a_odd}
\end {eqnarray}
For $n=0$, Eq.~(\ref {a_even}) is replaced by $a_0 = -\pi b_0 \,
\tilde c_{-1}$. Note that the absorptive coefficient at any given
order is determined by the dispersive coefficients at lower orders.

Conversely, expanding
\begin {equation}
    C(z) = A(z) / \sin (\pi b_0 z) \,
\label {C=A/sin}
\end {equation}
yields the inverse relations,
\begin {eqnarray}
    n \, c_n &=& \sum_{k=0}^{[ n/2 ]} \> {n! \over (2k)! \> (n{-}2k)!}
\; |(2^{2k}{-}2) B_{2k}| \> (\pi b_0)^{2k-1} \, a_{n-2k+1} \,,
\label {nc_n}
\end {eqnarray}
and $\tilde c_0 = a_1 / (\pi b_0)$.  Here $[x]$ denotes the integer
part of $x$, and $B_n$ are the Bernoulli numbers. Note that the
dispersive coefficient at a given order is determined by the
absorptive coefficient at one higher order plus absorptive
coefficients at lower orders.

    The existence of zeros in the $\sin (\pi b_0 z)$ denominator in
the relation between Borel transforms (\ref {C=A/sin}) implies that
the Borel transform of the dispersive coefficients $C(z)$ {\em must\/}
have singularities at integer multiples of $1/b_0$ unless the
absorptive transform $A(z)$ has compensating zeros.%
\footnote
    {%
 From the definition (\ref {1-term A(z)}), one sees that $A(z)$
does have a zero at the origin.  Hence $z = 0$ is not a singularity of
$C(z)$.  } These singularities constrain the possible large order
behavior of the dispersive perturbative coefficients \cite
{Brown&Yaffe}.  This is discussed in greater generality below.

\subsection {General beta functions and renormalization schemes}

    In the body of this paper, we show that the preceding one-term
beta function results do have a natural generalization to the case of
a general beta function.  The exact form of the results depends on the
coefficients (\ref {beta-function-expansion}) of the beta function
which in turn depend on the scheme used for defining the renormalized
coupling.  As is well known and will be reviewed at the beginning of
Sec.~V, the first two coefficients in the expansion of the beta
function are independent of the choice of scheme, while all higher
coefficients may be arbitrarily adjusted by a redefinition of the
renormalized coupling \cite {tHooft,Gross}.  We shall find it
convenient to choose a definition of the renormalized coupling for
which the {\em inverse\/} beta function contains only two terms,%
\begin {eqnarray}
	    {1 \over \beta (g^2)}
	    &=&\,
	    -{1 \over b_0 g^4} +
{\lambda \over b_0 g^2} \,,
\label{recycle}\\
\noalign {\hbox {or}}
	    \beta (g^2)
	    &=&
	    -b_0 g^4 \bigm/ (1 - \lambda g^2) \,,
\\
\noalign {\hbox {where}}
	    \lambda
	    &\equiv&
	    b_1 / b_0 \,.
\end {eqnarray}
With this choice, the solution of the renormalization group equation
(\ref {beta-function}) is particularly simple,
\begin {equation}
	{q^2 \over \mu^2} =
	\left( {g^2(q^2) \over g^2(\mu^2)}
    \right)^{\lambda/b_0} \exp \left\{ {1 \over b_0 g^2(q^2)}
- {1 \over b_0 g^2(\mu^2)} \right\} \,,
\label {t/mu^2}
\end {equation}
or
\begin{equation}
q^2 = \Lambda^2 \,
\left( g^2(q^2) \right)^{\lambda/b_0} \exp
\left\{ {1 \over b_0 g^2(q^2)} \right\} \,,
\label{qsquare}
\end {equation}
where
\begin{equation}
\Lambda^2
    \equiv \mu^2 \,
    \left( g^2(\mu^2) \right)^{-\lambda/b_0} \exp
\left\{- {1 \over b_0 g^2(\mu^2)} \right\}
\label{MASS}
\end {equation}
is the physical, renormalization group invariant mass scale of the
theory.

This choice of the beta function will also greatly reduce the algebra
in our work. Another simple alternative is to choose a different
coupling $\bar g^2$ whose beta function has only two terms
\cite{tHooft},
\begin{equation}
	\bar \beta (\bar g^2) \equiv \mu^2 \, d\bar g^2 / d\mu^2
       = -b_0 \bar g^4 \, (1 + \lambda \bar g^2) \,.
\end{equation}
As will be shown at the beginning of Sec.~V, the two schemes are
related by a shift in the inverse coupling, $ \bar g^{-2} = g^{-2} -
\lambda $. Moreover, as is also shown in Sec.~V, such a redefinition
only changes the Borel transforms in the two schemes by a simple
exponential factor,
\begin{equation}
	\bar A(z) = e^{- \lambda z} \, A(z) \,, \qquad
	\bar C(z) = e^{- \lambda z} \, C(z) \,.
\label{zhai}
\end{equation}
Thus our results employing the inverse two-term beta function may be
easily converted to the other case where the beta function itself has
only two terms.

\subsection {Modified Borel transforms}

    Remarkably, the simple relation (\ref {A=sin C}) between Borel
transforms with a one-term beta function does generalize to a two-term
(inverse) beta function provided that one considers a suitably
``modified'' Borel transform.  As shown in Sec.~\ref
{two-term-results}, one is naturally led to introduce the following
definition.  Given an asymptotic series
\begin {equation}
    f(z) \sim \sum_{n=0}^\infty f_n \> z^n \,,
\end {equation}
the {\em modified\/} Borel transform is defined as
\begin {equation}
    {\cal F}(z) \equiv \sum_{n=0}^\infty \>
    {\Gamma(1{+}\lambda z)
    \over \Gamma(n{+}1{+}\lambda z)} \; f_n \> z^n \,.
\label{Yaffe}
\end {equation}
Just like the ordinary Borel transform, this modified transform may be
viewed as a generating function for the coefficients $\{ f_n \}$.
Provided that the coefficients $\{ f_n \}$ grow no faster than $n! \>
k^n$ (for some constant $k$), ${\cal F}(z)$ is analytic in a
neighborhood of the origin.  The coefficients $\{ f_n \}$ may be
extracted from the derivatives of the transform ${\cal F}(z)$
evaluated at the origin.  Because of the presence of the shift by
$\lambda z$ in the argument of the gamma functions, a given
coefficient $f_n$ is not simply proportional to the $n$-th derivative
of $\cal F$.  Instead, $f_n$ is given by a linear combination of the
first $n$ derivatives of ${\cal F}(z)$ evaluated at the origin.  In
Sec.~V [c.f. Eq.~(\ref{stupid})] we derive the contour integral
representation
\begin {equation}
    f_n = \oint {dz \over 2\pi i z} \> \chi_n (1/z) \, {\cal F}(z) \,,
\label {F->f_n}
\end {equation}
where $\chi_n (y)$ is the $n$-th order polynomial
\begin {equation}
    \chi_n(y) = \delta_{n,0} +
    n \> y^n \, {\Gamma (n{+}\lambda/y) \over \Gamma (1{+}\lambda/y)}
    \,,
\label {chi_n}
\end {equation}
and the contour circles the origin.  The residue of the integrand in
Eq.~(\ref {F->f_n}) generates the appropriate linear combination of
derivatives.

    The modified Borel transform is related to the ordinary Borel
transform
\begin {equation}
    F(z) \equiv \sum_{n=0}^\infty \> {f_n \over n!} \; z^n
\end {equation}
through the integral relation
\begin {equation}
    F(z) = \oint {dy \over 2\pi i y} \>
 \left(
1 + {z/y \over (1 {-} z/y)^{1+\lambda y}}
 \right) 	 {\cal
F}(y) \,,
\label {cal F -> F}
\end {equation}
where the contour wraps about the cut connecting the branch-points of
the integrand at $y=0$ and $y=z$ and excludes any singularities of
${\cal F}(y)$.  This result follows directly from Eqs.~(\ref {F->f_n})
and (\ref {chi_n}) by using the generalized binomial theorem. An
alternative derivation is given in Sec.~V [c.f. Eq.~(\ref{INV})].  It
will be shown in Sec.~V [Eq.~(\ref{intrep})] that the inverse of this
relation also has a simple form,
\begin {equation}
  {\cal F}(z) = \lambda \int_0^z dy \; (1-y/z)^{\lambda z -1} F(y) \,.
\label {F -> cal F}
\end {equation}
 The integral relation (\ref {cal F -> F}) shows that the domain of
analyticity of the ordinary Borel transform includes that of the
modified Borel transform\footnote{ Actually, $F(z)$ is analytic in a
larger domain than ${\cal F}(z)$ since the overall factor of $\Gamma
(1+\lambda z)$ in the definition of the modified Borel transform
(\ref{Yaffe}) causes ${\cal F}(z)$ to have simple poles at $z =
-n/\lambda , \; n = 1,\, 2,\, \ldots$. These poles are removed by the
integral transform (\ref{cal F -> F}) which yields the standard
transform $F(z)$. The overall factor of $\Gamma ( 1+ \lambda z)$
could, of course, be omitted in the definition of the modified
transform, but it will prove convenient not to do so.} (since
singularities in $F(z)$ only develop when the contour is pinched
between the branch point at $z$ and a singularity of ${\cal F}(z)$).
In Appendix \ref {Borel asymptotics} it is shown that a basic effect
of the transformation (\ref {cal F -> F}) or its inverse (\ref{F ->
cal F}) is to shift the exponents of algebraic singularities. If the
modified transform has the singular behavior
\begin {eqnarray}
    {\cal F}(z) &\sim& (1-(z/R))^{-\alpha} \,,
\label {mod F sing}
\\
\noalign {\hbox {then the standard Borel transform has the singular
behavior }} F(z) &\sim& (1-(z/R))^{-\alpha - \lambda R} \;
{
 \Gamma (\alpha {+} \lambda R) \over
\Gamma (\alpha) \,
 \Gamma (1 {+} \lambda R)
}
\label {mod F sing -> F sing}
\end {eqnarray}
as $z \to R$, and conversely. The sub-leading, non-analytic terms in
this correspondence are suppressed by a relative factor of
$(1{-}(z/R)) \, \ln (1{-}(z/R))$.

    Just as for the usual Borel transform, the location and nature of
the singularities of ${\cal F}(z)$ closest to the origin determine the
leading asymptotic behavior of the coefficients $\{ f_n \}$.  If the
modified transform ${\cal F}(z)$ has a radius of convergence larger
than $K$, then a trivial bound of the integral (\ref {F->f_n}) using
the asymptotic form of the gamma function shows that the large-order
the growth of the coefficients $\{ f_n \}$ is bounded by
\begin {equation}
    |f_n| \le C \, n! \, K^{-n} \, n^{|\lambda|K}
\end {equation}
for some constant $C$.  If the nearest singularity to the origin has
the form (\ref {mod F sing}), then the generalized binomial expansion
of the corresponding Borel transform (\ref {mod F sing -> F sing})
shows that the coefficients $\{ f_n \}$ have the large order behavior
\begin {equation}
    f_n \sim R^{-n} \> {\Gamma (n {+} \alpha {+} \lambda R) \over
\Gamma (\alpha) \, \Gamma(1 {+} \lambda R)} \,,
\end{equation}
with corrections suppressed by $ \ln n / n$, or, using the asymptotic
behavior of the gamma function,
\begin{equation}
    f_n \sim { n! \> R^{-n} \> n^{\alpha + \lambda R -1}
 \over
\Gamma (\alpha) \, \Gamma (1{+}\lambda R)}
^{\phantom
{\Big|}} .
\label {mod Borel asymptotics}
\end {equation}

    Given the Borel transform of an asymptotic series, one may
generate a function whose asymptotic expansion coincides with the
original series by performing a Laplace transform,
\begin {equation}
    f(y) \equiv {1 \over y} \int_0^\infty dz \; e^{-z/y} \> F(z) \,,
\label {inverse-Borel-a}
\end {equation}
since expanding $F(z)$ yields
\begin {equation}
    f(y) \sim \sum_{n=0}^\infty \> f_n \; y^n \,.
\end {equation}
In terms of the modified Borel transform, the same construction reads
\begin {equation}
    f(y) \equiv {1 \over y} (1 {-} \lambda y) \int_0^\infty dz \;
e^{-z/y} \>
{\; (z/y)^{\lambda z} \over \Gamma
(1{+}\lambda z)}
\; \> {\cal F}(z) \,.
\label {inverse-mod-Borel-a}
\end {equation}
This is derived in Appendix \ref {Borel asymptotics}.  If $f_0$
vanishes, then the function
\begin{equation}
    \bar f(y) \equiv \int_0^y dy' \, f(y') / y' \sim \sum_{n=1}^\infty
\, f_n \, y^n / n \,
\label{simpler}
\end{equation}
has a slightly simpler form.  Inserting
Eq.~(\ref{inverse-mod-Borel-a}) into the integral in
Eq.~(\ref{simpler}) and using
\begin{equation}
    {1 \over y'^2}(1 - \lambda y') e^{-z/y'}
\left ( z/y'
\right )^{\lambda z}
= {1 \over z} {\partial \over
\partial y'} \left \{ e^{-z/y'}
\left (z/y'
\right)^{\lambda z}
\right \}
\end{equation}
yields
\begin {equation}
    \bar f(y) = \int_0^\infty {dz \over z} \; e^{-z/y} \>
{ (z/y)^{\lambda z} \over \Gamma (1{+}\lambda z) }
\>
{\cal F}(z) \,.
\label {inverse-mod-Borel-b}
\end {equation}
This will be the appropriate ``inverse'' to use if ${\cal F}(z)$ is
the modified Borel transform of a series $\{ n \, g_n \}$ where each
coefficient is scaled by $n$, since in this case $\bar f(y) \sim
\sum_{n=1}^\infty \, g_n \, y^n \,$.  This form will be used below.%

If the (ordinary or modified) Borel transform is analytic in a
neighborhood of the real axis and is well behaved as $z \to \infty$,
then the inverse transform (\ref {inverse-Borel-a}) or (\ref
{inverse-mod-Borel-a}) defines a unique ``sum'' of the asymptotic
series which satisfies certain boundedness and analyticity conditions
\cite {Hardy}.
If the Borel transform has a singularity at some point $z_0$ on the
positive real axis, then different contour prescriptions for the
integrals (\ref {inverse-Borel-a}) or (\ref {inverse-mod-Borel-a})
will produce ``sums'' of the asymptotic series $\sum_n f_n \, y^n$
which differ by exponentially small terms of order $\exp (-z_0/y)$.
This will be discussed in more detail below.

\subsection {General beta function results}

Let ${\cal A}(z)$ denote the modified Borel transform of the
absorptive coefficients $\{ n \, a_n \}$,
\begin {equation}
    {\cal A}(z) =
\sum_{n=1}^\infty \> 	{\Gamma(1{+}\lambda z)
\over \Gamma(n{+}1{+}\lambda z)} \;
n \, a_n \> z^n \,.
\label {A(z)}
\end {equation}
Similarly, let ${\cal C}(z)$ denote the modified transform of the
dispersive coefficients $\{ n \, c_n \}$ with a suitably chosen
constant piece,
\begin {equation}
    {\cal C}(z) =
(\tilde c_0 - \lambda \tilde c_{-1}) +
\sum_{n=1}^\infty \>
{\Gamma(1{+}\lambda z) \over
\Gamma(n{+}1{+}\lambda z)} \;
n \, c_n \> z^n \,.
\label {C(z)}
\end {equation}
In Sec.~\ref {two-term-results} we show that these transforms obey the
same simple relation as in the one-term beta function case,
\begin {equation}
    {\cal A}(z) = \sin (\pi b_0 z) \, {\cal C}(z) \,.\qquad
\label {mod A=sin C}
\end {equation}

    When Eq.~(\ref {F->f_n}) is used to extract the original
absorptive coefficients from this relation, one finds that%
\footnote
    {%
 Amusingly, the explicit relation (\ref {a_n = sum c_n}) between
the absorptive and dispersive coefficients, and its inverse (\ref {c_n
= sum a_n}), were derived before the appropriate definition of the
modified Borel transform satisfying (\ref {mod A=sin C}) was found.
Direct (but tedious) methods for obtaining these results are described
in Appendix \ref {Alternate derivations}.  }
\begin {eqnarray}
	a_n &=&
(\tilde c_0 {-} \lambda \tilde c_{-1}) \>
\sum_{k=0}^{[{n-1 \over 2}]} \>
(-)^k {(\pi b_0)^ {2k+1} \over
(2k+1)!} \>
\lambda^{n-2k-1} \> I_{0,2k}^{n-1} \nonumber
\\
    &&{} +
\sum_{m=1}^{n-1} m \> c_m 	\sum_{k=0}^{[{n-m-1
\over 2}]}
(-)^k \> {(\pi b_0)^{2k+1} \over (2k+1)!} \>
\lambda^{n-m-2k-1} \> I_{m,2k}^{n-1}\,.
\label {a_n = sum c_n}
\end {eqnarray}
Here the $\{ I^n_{m,l} \}$ are combinatorial factors defined by the
generating function
\begin {equation}
    {\Gamma(n+1+x) \over \Gamma(m+1+x)}
\equiv
\sum_{l=0}^{n-m} x^{n-m-l} I_{m,l}^n \,.
\label {I^n_ml}
\end {equation}
Alternatively, applying Eq.~(\ref {F->f_n}) to
\begin {equation}
    {\cal C}(z) = {\cal A}(z) / \sin (\pi b_0 z) \,
\label {mod C=A/sin}
\end {equation}
yields the inverse relations
\begin {equation}
     c_n = {a_{n+1} - \lambda a_n \over \pi b_0 \, n } +
	\sum_{m=1}^{n-1} \> m \, a_m
	\sum_{k=1}^{[{n-m+1\over 2}]}
	{|(2^{2k} {-} 2) B_{2k}| \over (2k)!} \>
	(\pi b_0)^{2k-1} \>
	\lambda^{n-m-2k+1} \> I_{m,2k-2}^{n-1} \,,
\label {c_n = sum a_n}
\end {equation}
together with $\tilde c_0 = (a_1 {-} \lambda a_0) / (\pi b_0)$ and
$\tilde c_{-1} = -a_0 / (\pi b_0)$.  These results are derived in
Sec.~\ref {two-term-results}. It is easy to verify that they reduce to
the previous results (\ref{a_even}), (\ref{a_odd}), and (\ref{nc_n})
for the case of the one-term beta function when the limit $\lambda \to
0$ is taken. Note that, just as in the case of the one-term beta
function, the relations (\ref{a_n = sum c_n}) and (\ref{c_n = sum
a_n}) have a ``triangular structure'': The right-hand side of
Eq.~(\ref{a_n = sum c_n}) involves coefficients of lower order; the
right-hand side of Eq.~(\ref{c_n = sum a_n}) involves a coefficient of
one higher order plus coefficients of lower order.

\subsection {Large order behavior of the dispersive coefficients}

    The existence of zeros in the $\sin (\pi b_0 z)$ denominator in
the relation between the modified Borel transforms (\ref {mod
C=A/sin}) implies that ${\cal C}(z)$ will have singularities at all
non-zero integer values of $b_0 \, z$ unless ${\cal A}(z)$ has
compensating zeros.  Hence, one of the following possibilities for the
large-order behavior must occur:
\begin {enumerate}
    \dimen0 = \textwidth \advance\dimen0 by -\leftmargin
    \item
    If ${\cal A}(z)$ has a radius of convergence greater than $1/b_0\,$
    (so that the absorptive coefficients $\{ a_n \}$ grow slower than
    $n! \, K^n$ for some $K < 1/b_0$),
    then ${\cal C}(z)$ will have simple poles at $z = \pm 1 / b_0$.
    If the residues ${\cal A}_\pm \equiv {\cal A}(\pm 1 / b_0)$
    are not both zero,
    Eqs.~(\ref{mod F sing}) and (\ref {mod Borel asymptotics})
    (for $\alpha =1$) show that
	the dispersive coefficients will have large-order
    behavior which is completely determined by these residues,%
    {%
    \advance\hsize by -\leftmargin
    \begin {equation}
	c_n \sim \> (n{-}1)!
	\left\{
	    {
		{\cal A}_+ \> b_0^n \> n^{\lambda/b_0} \over
		\pi \, \Gamma (1 {+} \lambda/b_0)
	    }
	-
	    {
		{\cal A}_- \> (-b_0)^n \> n^{-\lambda/b_0} \over
		\pi \, \Gamma (1 {-} \lambda/b_0)
	    }
	\right\}
    \label{c_n prepred}
    \end {equation}%
    }%
    as $n \to \infty$.
   \item
   If ${\cal A}(z)$ has a radius
   of convergence equal to $1/b_0$,
   so will ${\cal C}(z)$. If the
   singularity nearest to the orgin
   lies on the real axis, then
   the dispersive coefficients will grow faster than the absorptive
   coefficients by a single power of
   $n$.\footnote{ If ${\cal A}(z)$ has
   both a non-zero value and a branch point at $b_0 z = \pm 1$ with a
   vanishing discontinuity at the branch point [e.g., ${\cal A}(z) \sim
   {\cal A}_+ + (1 - b_0 z)^\alpha$, with $\alpha > 0$], then the
   dominant singularity in ${\cal C}(z)$ is a simple pole at $b_0 = \pm
   1$, and the large order behavior of the dispersive coefficients is
   governed by Eq.~(\ref{c_n prepred}). In this case, the dispersive
   coefficients grow faster than the absorptive coefficients by more than
   one power of $n$.}
   For example, if the absorptive coefficients
   behave for large $n$ as
   {%
   \advance\hsize by -\leftmargin
\begin {equation}
  a_n \sim \> (n{-}1)!
   \left\{
  {{\cal A}_+ \> b_0^n \> n^{\lambda/b_0 + \gamma_+}
  \over \Gamma (1 {+} \lambda/b_0)}
   + 	{{\cal A}_- \>
  (-b_0)^n \> n^{- \lambda/b_0 + \gamma_-}
  \over \Gamma
  (1 {-} \lambda/b_0)}
  \right\}
\label {a_n pred}
\end{equation}%
  }%
  for some constants ${\cal A}_\pm$ and
  $\gamma_\pm > -1$,
  then from Eq.~(\ref {mod Borel asymptotics})
  the modified Borel transform ${\cal A}(z)$ will have the
  behavior
$  {\cal A}(z) \sim {\cal A}_\pm \, 	 (1
   \mp b_0 z)^{-1-\gamma_\pm} \,
   \Gamma (1{+}\gamma_\pm) 	$
   as $b_0 z \to \pm 1$.
   Dividing by $\sin(\pi b_0 z)$ gives the
   behavior of the modified dispersive transform,
	$ 	 {\cal
C}(z) \sim \pm {\cal A}_\pm \,
 (1 \mp b_0 z)^{-2-\gamma_\pm} \,
\Gamma (1{+}\gamma_\pm)
 / \pi 	$.
  Thus the dispersive coefficients will grow like
{%
\advance\hsize by -\leftmargin
\begin {equation}
c_n \sim \> n!   \left\{
{{\cal A}_+ \> b_0^n \> n^{ \lambda/b_0 + \gamma_+}
\over
\pi \, (1 {+} \gamma_+) \, \Gamma (1 {+} \lambda/b_0)}
 -
{{\cal A}_- \> (-b_0)^n \> n^{ - \lambda/b_0 + \gamma_-}
\over \pi \, (1 {+} \gamma_-) \, \Gamma (1 {-} \lambda/b_0)}
\right\}
\label {c_n pred}
\end {equation}%
}%
   as $n \to \infty$.
   \item
   If ${\cal A}(z)$ has a radius of
   convergence less than $1/b_0$,
   so will ${\cal C}(z)$.  	In
   this case, the dispersive coefficients will have the same large
  order behavior (within an overall constant factor) 	as the
  absorptive coefficients, with both growing faster than $b_0^n
  \, n!$ as $n \to \infty$.
\end {enumerate}

    These conditions on the possible large-order behavior are
independent of the specific dynamics of the asymptotically free theory
and follow solely from the existence of renormalized perturbation
theory.  Each of the possible behaviors above is fully consistent with
the constraints of analyticity and the renormalization group,
contradicting the claim of unique large order behavior asserted in
\cite {West}. (The asymptotic behavior given in this reference also
does not obey the constraint given in point 3 above, as shown in
Appendix A.) However, in view of Eqs.~(\ref{c_n prepred}), (\ref{c_n
pred}) and the asymptotic form of the gamma function, large-order
behavior of the form
\begin {equation}
    c_n \sim \>
{\cal C}_+ \> b_0^n \> \Gamma (n+\gamma_+) -
{\cal C}_- \> (-b_0)^n \> \Gamma (n-\gamma_-)
\label {generic-prediction}
\end {equation}
for the dispersive coefficients is, in some measure, a {\em generic\/}
possibility.  This behavior (for some values of ${\cal C}_\pm$ and
$\gamma_\pm$) will occur {\em unless\/} either ({\em i}) the
absorptive transform ${\cal A}(z)$ is singular within $b_0 |z| \le 1$
(in which case both absorptive and dispersive coefficients grow more
rapidly than (\ref {generic-prediction})), or ({\em ii}) the
absorptive transform has zeros at both $z = \pm 1/b_0$ and a radius of
convergence greater than $1/b_0$ (in which case the coefficients grow
more slowly than (\ref {generic-prediction})).

\subsection {Renormalons and operator product expansions}

Explicit studies of perturbation theory in QCD show that \cite
{tHooft,Zinn-Justin,Parisi,Mueller}:
\begin {itemize}
 \item [({\it i\/})]
The ultraviolet behavior of individual
multi-loop diagrams
can generate contributions behaving as
$c_{m+1} \sim (-b_0/k)^m \> m!$, for $k = 1, 2, \ldots \,$,
leading to singularities in the Borel transform $C(z)$
at the points
$z = -k/b_0$ on the negative real axis.
Near the first
singularity \cite {Parisi},
$ 	 C(z) \sim (b_0 \, z {+}
1)^{-1+\gamma}
$, 	where $\gamma$ is related to the anomalous
dimension of
local operators of dimension six.  	These
contributions are referred to as {\em ultraviolet renormalons}.  \item
[({\it ii\/})]
The infrared behavior of multi-loop diagrams 	can
generate contributions behaving as
$c_{m+1} \sim (b_0/k)^m \>
m!$, for $k = 2, 3, \ldots \,$,
corresponding to singularities
in ${\cal C}(z)$ at the points
$z = k/b_0$ on the positive real axis.
Near the first singularity \cite {Mueller},
$ 	 C(z) \sim
(b_0 \, z {-} 2)^{-1-2 \lambda /b_0} \,
$, 	or
equivalently (using (\ref {F -> cal F}))
the modified transform
${\cal C}(z)$ has a simple pole, with a
subleading logarithmic
branch cut.\footnote{This entails a branch point in ${\cal A}(z)$ with
a vanishing discontinuity as discussed in the previous footnote.}
These contributions are referred to as {\em infrared
renormalons}.  \item [({\it iii\/})]
Instanton--anti-instanton
pairs generate singularities in the Borel transform
on the
positive real axis (starting at $z = 16\pi^2$)
to the right of the
leading infrared renormalon singularity.  \item [({\it iv\/})]
No
other sources of singularities in the Borel transform are known.
\end {itemize}

    The presence of infrared renormalon singularities on the positive
real axis is directly related to the existence of non-perturbative
vacuum expectation values of composite operators \cite {David}.  The
operator product expansion of two electromagnetic currents reads
\begin {equation}
i \, {\cal T} \left( j_\mu (y + {\textstyle {1 \over 2}} x) \,
	j_\nu (y - {\textstyle {1 \over 2}} x)
\right) \sim \sum_i \> C_{\mu\nu}^i (x) \; {\cal O}_i(y) \,,
\label {OPE}
\end {equation}
where $\{ {\cal O}_i \}$ denotes the appropriate set of local
operators.  The Fourier transform of the vacuum expectation value of
this expansion gives
\begin {equation}
    K(-q^2) \sim
\sum_i \> 	\tilde C^i (q) \> \langle 0|
{\cal O}_i |0 \rangle \,,
\end {equation}
where
\begin {equation}
    \tilde C^i (q) \equiv {g^{\mu\nu} \over 3 q^2} \int d^4x \>
e^{-iqx} \; C_{\mu\nu}^i (x) \,.
\end {equation}
The coefficient functions for the scalar operators $\{ {\cal O}_i \}$
with non-vanishing vacuum expectation values have the form
\begin{equation}
\tilde C^i(q) = \left( q^2 \right)^{-d_i/2}
D^i\left( q^2/\mu^2 ,\, g^2(\mu^2) \right) \,,
\end{equation}
where the $d_i$ are the physical dimensions of the operators ${\cal
O}_i$ and the $D^i$ are dimensionless functions which have a
perturbative expansion in powers of $g^2(\mu^2)$.  For QCD, the lowest
dimension gauge-invariant composite operators with non-vanishing
vacuum expectation values are the unit operator $\hat 1$, with
dimension $d = 0$, $[F^{\mu\nu} F_{\mu\nu}]$ and $ [m \, \bar\psi
\psi]$, both of dimension $d = 4$, followed by various operators of
dimension 6, 8, $\ldots\,$.\footnote{ Chiral symmetry requires that
non-chirally invariant operators like $[\bar \psi \psi]$ be
accompanied by a factor of the quark mass.  Hence, only even dimension
composite operators appear in the expansion.} The entire asymptotic
expansion of the Euclidean correlation function (\ref {K-exp}) is
contained in the coefficient function of the unit operator.  The
dimensionless functions which are not associated with the unit
operator obey a homogeneous renormalization group equation
\begin{equation}
\left\{ \mu^2 { \partial \over \partial \mu^2 }
+ \beta(g^2) {\partial \over \partial g^2} -\gamma^i(g^2) \right\} D^i
\left( q^2/\mu^2 ,\, g^2(\mu^2) \right) = 0 \,,
\end{equation}
where the $\gamma^i$ are the anomalous dimensions of the operators
${\cal O}^i$. The renormalization group may be used to transfer the
momentum dependence into the running coupling,
\begin{equation}
D^i \left( q^2/\mu^2 ,\, g^2(\mu^2) \right) = \exp \left\{ -
\int_{g^2(\mu^2)}^{g^2(q^2)} dg^2 {\gamma^i(g^2)
\over \beta(g^2) } \right\} D^i \left( 1,\, g^2(q^2) \right) \,.
\end{equation}
By using the expansion (\ref{beta-function-expansion}) for the beta
function and writing $\gamma^i(g^2) = \gamma_0^i \> g^2 + \gamma^i_1
\> g^4 + \ldots $ one finds that
\begin{equation}
- \int_{g^2(\mu^2)}^{g^2(q^2)} dg^2 {\gamma^i(g^2)
\over \beta(g^2) } = {\gamma_0^i \over b_0 } \ln {g^2(q^2) \over
g^2(\mu^2) } + \cdots \,,
\end{equation}
where the ellipsis stands for a power series in $g^2(q^2)$ minus the
same series in $g^2(\mu^2)$. The terms involving $g^2(\mu^2)$ may be
absorbed by making a suitable multiplicative redefinition of the
renormalized operator ${\cal O}_i$ to yield a renormalization group
invariant which is independent of $\mu^2$. The vacuum expectation
value of this invariant produces a dimensionless numerical constant
times $\Lambda^{d_i}$, where $\Lambda$ is the renormalization group
invariant mass parameter defined in Eq.~(\ref{MASS}). The power series
in $g^2(q^2)$ can be absorbed into a redefinition of the coefficient
function $D^i\left( 1,\, g^2(q^2) \right)$. Thus, a given term in the
operator product expansion produces a contribution to the current
correlation function of the form
\begin{equation}
K^i(-q^2) \sim \left( {\Lambda^2 \over q^2 } \right)^{d_i/2} \left(
g^2(q^2) \right)^{\gamma_0^i / b_0} \bar D^i \left( g^2(q^2) \right)
\,,
\end{equation}
where $\bar D^i \left( g^2(q^2) \right) $ admits an asymptotic
expansion in powers of $g^2(q^2)$. Expressing $q^2 / \Lambda^2$ in
terms of the running coupling defined by the two-term inverse beta
function [ Eq.~(\ref{t/mu^2}) ] shows that these contributions give
essential singularities at the origin in the $g^2(q^2)$ plane. The
leading behavior at $g^2(q^2) = 0$ for each term in the operator
product expansion is given by
\begin{equation}
K^i(-q^2) \sim \left( g^2(q^2) \right)^{ (\gamma_0^i -
\lambda d_i/2 ) / b_0} \exp \left\{ - {d_i \over 2 b_0 g^2(q^2) } \right\}
 \tilde D^i \,,
\label{higher-dim-contrib}
\end{equation}
where $ \tilde D^i \equiv \bar D^i(0) $ is a constant.

 The contribution from a composite operator is inherently scheme
dependent; one can always redefine the operator by adding a constant
multiple of the unit operator.  This changes the vacuum expectation
value of the operator, at the cost of moving (part of) the
non-perturbative contribution (\ref {higher-dim-contrib}) into the
coefficient function of the unit operator.  Consequently, in any
method to ``resum'' the perturbation series, one should expect to find
ambiguities of precisely the form (\ref {higher-dim-contrib}).
Infrared renormalon singularities in the Borel transform are precisely
the reflection of these ambiguities.  As mentioned earlier, each
singularity on the positive real axis generates a non-perturbative
ambiguity in the inverse Borel transform.  It is instructive to work
out a specific example. We consider the case where the modified Borel
transform ${\cal C}(z)$ has a branch point at $b_0 z = d$ near which
${\cal C}(z) \sim ( d - b_0 z)^{\sigma -1}$. A singularity of this
form in the inverse transform (\ref{inverse-mod-Borel-b}) creates an
ambiguity resulting from different possible choices for routing the
contour about the branch cut which starts at $z = d/b_0$. The
discontinuity across this cut gives a measure of this ambiguous
contribution. Retaining only the leading term near $g^2(q^2) = 0$ we
obtain
\begin{eqnarray}
\Delta K(-q^2) &\sim& {\rm const.} \int_{d/b_0}^\infty dz \exp\left\{ - {z
\over g^2(q^2)} \right\} \left( g^2(q^2) \right)^{-\lambda z}
(b_0 z - d )^{\sigma -1}
\nonumber\\
		&\sim& {\rm const.'} \left( g^2(q^2) \right)^{\sigma -
\lambda d / b_0} \exp \left\{ - {d \over b_0 g^2(q^2) } \right\} \,.
\end{eqnarray}
This is precisely the structure of a non-perturbative operator-product
contribution (\ref{higher-dim-contrib}) if we identify $d = d_i/2$ and
$\sigma = \gamma_0^i / b_0$.

    With one exception, the singularities in the modified transform
${\cal C}(z)$ generated by the zeros of the $\sin (\pi b_0 z)$
denominator in Eq.~(\ref{mod C=A/sin}) are at precisely those
locations which correspond to ultraviolet and infrared renormalons.
The exception is the zero at $b_0 z = 1$.  A singularity at this
location will generate a non-perturbative ambiguity of order $1/q^2$
in the inverse Borel transform which, because there is no gauge
invariant local scalar operator of dimension 2, cannot be attributed
to any physical vacuum expectation value.  The absence of this
singularity is possible only if the modified absorptive Borel
transform ${\cal A}(z)$ has a zero at $b_0 \, z = 1$.  This constraint
has not been previously noted.  Alternatively, if a singularity at
$b_0 z = 1$ does exist, leading to $O(\Lambda^2 /q^2)$ contributions
to the inverse Borel transform, then this would have to be interpreted
as an unexpected non-perturbative correction to the coefficient
function of the unit operator in the operator product
expansion.\footnote{ In massless QCD, any $1/q^2$ correction must have
a non-perturbative origin. However, with non-vanishing quark masses in
the Lagrangian, $ O(m^2 /q^2)$ terms, calculable in perturbation
theory, do appear in the operator product expansion.} This would be a
major problem for phenomenological applications of the operator
product expansion ({\em e.g.}, QCD sum rules \cite {QCD-sum-rules})
which are based on the assumption that there exists a range of momenta
where non-perturbative $\langle [F^2] \rangle/q^4$ and $\langle [m
\bar\psi\psi] \rangle/q^4$ contributions are significant and correctly
parameterize the leading non-perturbative effects, while
simultaneously all coefficient functions may be well approximated by
the first term or two of their perturbative expansions.  No convincing
argument demonstrating either the presence or absence of such $1/q^2$
corrections is known to the authors.%
\footnote
    {%
    Conceivably, similar $1/q^2$ non-perturbative contributions
might appear in many other correlation functions.  Such contributions
may already have been seen in Wilson loop expectation values, where
available numerical data suggests the presence of corrections
proportional to the area of the loop in the limit of {\em small\/}
loops (%
{\em i.e.}, for loops large compared to the lattice spacing
but small compared to $1/\Lambda$) \cite {Lepage}.  }

\section {Renormalization Group and Analyticity}
\label {ren group}

  We turn at last to the details of our work, beginning with a review
of the renormalization group equation for the current-current
correlation function and the origin of the large momentum asymptotic
expansion (\ref {K-exp}).  Using a mass-independent renormalization
scheme, the dimensionless function $K(t)$ depends on the
renormalization point $\mu^2$, the renormalized coupling $g^2(\mu^2)$,
and any mass parameters $m(\mu^2)$ in the form
\begin {equation}
    K(t) = K(t/\mu^2, g^2(\mu^2), m^2(\mu^2)/\mu^2 ) \,.
\label {K-dependence}
\end {equation}
The function $K(t)$ has a perturbative expansion in powers of
$g^2(\mu^2)$,
\begin {equation}
    K (t) \sim \sum_{n=0}^\infty \; k_n (t/\mu^2, m^2/\mu^2) \,
\left( g^2(\mu^2) \right)^n \,.
\label {pert_exp}
\end {equation}
Since all momentum dependence is hidden in the coefficients $k_n$,
this expansion is not directly useful for examining the large momentum
behavior of $K(t)$.  However, Eq.~(\ref {K-dependence}) implies that a
variation in $t$ is equivalent to a variation in $\mu^2$ combined with
a suitable compensating change in the coupling $g^2(\mu^2)$ and mass
$m(\mu^2)$.  The dependence of $K(t)$ on the renormalization point
$\mu$ is described by the inhomogeneous renormalization group equation
\begin {eqnarray}
    && \hspace* {-6ex} \mu^2 {d \over d\mu^2} \> K
(t/\mu^2,g^2(\mu^2),m^2(\mu^2)/\mu^2) = {}
\nonumber
\\
    && \biggl[
 \mu^2		 {\partial \over \partial \mu^2} +
\beta (g^2)
 {\partial \over \partial g^2} +
\delta (g^2) \, m^2 {\partial \over \partial m^2} \biggr] \,
 K
(t/\mu^2,g^2,m^2/\mu^2) =
D(g^2) \,,
\label {RGE}
\end {eqnarray}
where the beta function $\beta (g^2)$ describes the $\mu$-dependence
of the renormalized coupling,
\begin {equation}
    \mu^2 {d \over d \mu^2} \, g^2(\mu^2) = \beta (g^2(\mu^2)) \,,
\label {RGE-beta}
\end {equation}
and the anomalous dimension $\delta(g^2)$ characterizes the variation
of the running mass,
\begin {equation}
    \mu^2 {d \over d \mu^2}\, m^2(\mu^2) = m^2(\mu^2) \,\delta
(g^2(\mu^2))\,.
\label{deltarun}
\end {equation}
Because the electromagnetic currents are conserved they acquire no
anomalous dimension.  However, since the product of two current
operators is singular, one subtraction proportional to the unit
operator is required for the proper definition of the time-ordered
product in Eq.~(\ref {K_mu_nu}); the inhomogeneous term $D(g^2)$
characterizes the dependence of this subtraction term on the
renormalization scale.
The functions $D(g^2)$, $\beta(g^2)$, and
$\delta(g^2)$ have perturbative expansions of the form
\begin {eqnarray}
    D (g^2) &\sim& \> d_0 + d_1 \, g^2 + d_2 \, g^4 + \cdots \,,
\label {D-fcn-exp}
\\
    \beta (g^2) &\sim& -b_0 \, g^4 -b_1 \, g^6 - \cdots \,,
\label {beta-fcn-exp}
\\
    \delta (g^2) &\sim& \> \delta_0 \, g^2 +\delta_1 \, g^4 + \cdots
\,,
\label{delta-fcn-exp}
\end {eqnarray}
with $b_0 > 0$ in an asymptotically free theory like QCD.

    We will first consider the current-correlation function $K(t)$ in
the Euclidean region where $-t$ is real and positive.  To solve the
renormalization group equation (\ref {RGE}), one first introduces a
running coupling $g^2(-t)$ defined by
\begin {equation}
    \int_{g^2(\mu^2)}^{g^2(-t)} {d g^2 \over \beta(g^2) } \equiv \ln
\left( {-t \over \mu^2} \right) \,.
\label {running-coupling}
\end {equation}
The coupling $g^2(-t)$ is independent of the renormalization point
$\mu$ but obeys Eq.~(\ref {RGE-beta}) with $\mu^2$ replaced by $-t$.
Having defined the running coupling, we may now define the
momentum-dependent mass parameter
\begin{equation}
m^2(-t) = m^2(\mu^2) \exp\left\{ \int_{g^2(\mu^2)}^{g^2(-t)} dg^2 \,
{\delta(g^2) \over \beta(g^2)} \right\} \,.
\label{massrun}
\end{equation}
With these definitions in hand, the general solution of the full
renormalization group equation (\ref {RGE}) may be written as
\begin {equation}
	K ( t/\mu^2, g^2(\mu^2), m^2(\mu^2)/\mu^2 ) =
K (-1,
g^2(-t), m^2(-t)/(-t) ) -
\int_{g^2(\mu^2)}^{g^2(-t)} dg^2 \>
{D(g^2) \over \beta(g^2) } \,.
\label {K-rge-soln}
\end {equation}

This result may now be expanded in powers of $g^2(-t)$.  However, the
presence of the inhomogeneous term involving $D(g^2)$ alters the
perturbative expansion of $K(t)$.  To see this, we note that the
expansions (\ref {D-fcn-exp}) and (\ref {beta-fcn-exp}) imply that
\begin {equation}
    - \int_{g^2(\mu^2)}^{g^2(-t)} d g^2 {D(g^2) \over \beta(g^2) } =
{d_0 \over b_0}
 \left[ {1 \over g^2(\mu^2)} - {1 \over
g^2(-t)} \right]
+ \left( {d_1 \over b_0} - {d_0 b_1 \over
b_0^2} \right)
 \ln \left[ {g^2(-t) \over g^2(\mu^2)} \right] 	+
\cdots \,,
\label{inhomo}
\end {equation}
where the ellipsis stands for a power series in $g^2(-t)$ minus the
same series in $g^2(\mu^2)$.  All the terms involving $g^2(\mu^2)$ may
be absorbed in a single $\mu^2$-dependent parameter $\kappa (\mu^2)$.
The series in $g^2(-t)$ combines with the perturbative expansion (\ref
{pert_exp}) of $K(-1, g^2(-t), m^2(-t) / (-t))$ to yield a modified
expansion in powers of $g^2(-t)$.  Hence $K(t)$ has a large-$t$
asymptotic expansion of the form
\begin {equation}
    K(t) \sim \kappa (\mu^2)
+ \tilde c_{-1} \,
g^2(-t)^{-1}
+ \tilde c_0 \, \ln g^2(-t) 		+
\sum_{n=1}^\infty c_n \, g^2(-t)^n \,,
\label {K-exp-2}
\end {equation}
as quoted earlier in Eq.~(\ref {K-exp}).  The coefficients $\{ \tilde
c_{-1}, \tilde c_0, c_n \}$ are independent of $\mu^2$ and all
renormalized masses.  The mass independence follows since $K(t)$ is
finite in the massless limit and $m^2(-t)/t$ vanishes faster\footnote{
Inserting the expansions (\ref{beta-fcn-exp}) and
(\ref{delta-fcn-exp}) into the definition (\ref{massrun}) of the
momentum-dependent mass shows that, in the large $-t$ limit where
$g^2(-t)$ tends to zero,
$$
m^2(-t) \sim \left( g^2(-t) \right)^{-\delta_0/b_0} m^2 \,,
$$ where
$$
m^2 \equiv \left( g^2(\mu^2)
\right)^{\delta_0/b_0} \exp\left\{ -
\int_0^{g^2(\mu^2)} dg^2 \left[ {\delta(g^2) \over \beta(g^2)} + {\delta_0
\over b_0 g^2} \right] \right\} m^2(\mu^2) \,,
$$
is independent of the scale mass $\mu$. In QCD $\delta_0 < 0$, and
thus $m^2(-t)$ vanishes as a positive power of the coupling when $ -t
\to \infty$. The additional suppression by $1/t$ makes such mass terms
asymptotically insignificant.} than any power of $g^2(-t)$ as $t \to
\infty$.
All remaining $\mu$-dependence is contained in the
momentum-independent term $\kappa (\mu^2)$.  The presence of the
$1/g^2(-t)$ and $\ln g^2(-t)$ terms in this result may, at first
glance, appear odd.  However, since the expansion of
Eq.~(\ref{running-coupling}) gives
\begin{eqnarray}
	{1 \over g^2(-t)} &=&
{1 \over g^2(\mu^2)} + 	b_0 \ln \left(
{-t \over \mu^2} \right) -
{b_1 \over b_0} \ln \left[ {g^2(-t)
\over g^2(\mu^2) } \right] +
\cdots \,,
\label {coupling-expansion}
\end{eqnarray}
the $1/g^2(-t)$ term is precisely what is required to generate the
$\ln (q^2 / \mu^2)$ behavior of the free-field correlation function.
Similarly, the $\ln g^2(-t)$ term reflects the presence of $\ln
(\ln(q^2/\mu^2))$ terms in the large momentum behavior of $K(t)$.

It is worth noting that using different renormalization prescriptions
may result in the addition of a finite, $q^2$-independent term $
P(g^2(\mu^2))$ to the renormalized current correlation function
$K(t/\mu^2, g^2(\mu^2))$, with $P(g^2(\mu^2))$ having a power series
expansion
\begin{equation}
    P(g^2(\mu^2)) \sim \sum_{n=0}^{\infty} p_n \, g^{2n}(\mu^2) \,.
\end{equation}
Such a change will add a contribution
\begin{equation}
    \mu^2 {d P(g^2(\mu^2)) \over d \mu^2 }
= \beta (g^2(\mu^2))
{d P(g^2) \over d g^2 }
\label{PPP}
\end{equation}
to the inhomogeneous term $D(g^2(\mu^2))$ in the renormalization group
equation (\ref{RGE}). Since this shift in $D(g^2(\mu^2))$ has an
expansion in powers of $g^2(\mu^2)$ starting at order $ g^4(\mu^2)$,
we learn that, by appropriately choosing the renormalization scheme,
one can remove from $D(g^2)$ all but the first two terms in its
perturbative expansion, and obtain
\begin{equation}
    D(g^2) = d_0 + d_1 g^2\,.
\end{equation}
The remaining coefficients $d_0$ and $d_1$ cannot be altered by this
redefinition.  It is these terms in the function $D(g^2)$ that produce
the $1/g^2(-t)$ and $\ln g^2(-t)$ pieces in the asymptotic expansion
(\ref{K-exp-2}) of the correlation function $K(t)$.

The inhomogeneous term in the renormalization group equation for $
K(t/\mu^2, g^2(\mu^2)) $ can be avoided altogether if one studies
instead the differentiated function\footnote{ The occurrence of the
inhomogeneous term in the renormalization group is related to the need
for a subtraction in the Lehmann (dispersion relation) representation
for $ K(t/\mu^2, g^2(\mu^2)) $. The Lehmann representation for $t
\,\partial K(t/\mu^2, g^2(\mu^2)) /\partial t$ dispenses with this
additional subtraction constant.} $t \,\partial K(t/\mu^2, g^2(\mu^2))
/\partial t$.  Here, to keep the discussion simple, we consider only
the massless case.  Since this function satisfies a homogeneous
renormalization group equation, we may choose $\mu^2 = -t$ to obtain a
power series expansion in the running coupling $g^2(-t)$,
\begin{equation}
    t {\partial \over \partial t} K(t/\mu^2, g^2(\mu^2))
\sim
\sum_{n=0}^{\infty} k_n' \, g^{2n}(-t) \,.
\label{deriv}
\end{equation}
The expansion of the original correlation function may be recovered by
integrating over $t$ and using $dt/t = d g^2(-t) / \beta (g^2(-t))$,
\begin{equation}
    K(t/\mu^2, g^2(\mu^2)) \sim K(1, g^2(\mu^2))
+
\int_{g^2(\mu^2)}^{g^2(-t)} {d g^2 \over \beta(g^2)}
\,
\sum_{n=0}^{\infty} k_n' \, g^{2n} \,.
\label{away}
\end{equation}
Comparing this expansion with the previous results (\ref{inhomo}) and
(\ref{K-exp-2}) shows that the first two terms of the differentiated
current correlation function (and of the inhomogeneous term $D(g^2)$)
are given by
\begin{eqnarray}
	k_0' &=& - d_0 = b_0 \tilde c_{-1} \,,
\nonumber\\
	k_1' &=& -d_1 = b_1 \tilde c_{-1} - b_0 \tilde c_0 \,.
\label{mayday}
\end{eqnarray}

    Since the correlation function $K(t)$ is analytic throughout the
cut $t$-plane, in addition to the Euclidean space asymptotic expansion
(\ref {K-exp-2}), we may examine the asymptotic behavior of $K(t
e^{i\theta})$ as $t \to -\infty$ for an arbitrary phase $\theta$.  The
same steps as before yield
\begin {equation}
	K ( t e^{i\theta} /\mu^2, g^2(\mu^2), m^2(\mu^2)/\mu^2 ) =
K (-e^{i\theta}, g^2(-t), m^2(-t)/(-t) ) -
\int_{g^2(\mu^2)}^{g^2(-t)} dg^2 \>
{D(g^2) \over \beta(g^2) } \,,
\label {K-rge-soln-2}
\end {equation}
and the large-$t$ asymptotic expansion in powers of $g^2(-t)$ now
produces $\theta$-dependent (but $m$ and $\mu$ independent)
coefficients,
\begin {equation}
    K (t e^{i\theta}) \sim \kappa (\mu^2)
+ \tilde
c_{-1} \, i b_0 \theta
+ \tilde c_{-1} \, g^2(-t)^{-1}
+ \tilde c_0 \, \ln g^2(-t)
+ \sum_{n=1}^\infty
c_n(\theta) \> g^2(-t)^n \,.
\label {K-exp-3}
\end {equation}
Here, so as to simplify the later notation, we anticipate the result
that the phase dependence of the $g^2$ independent term has the simple
linear form $ \tilde c_{-1} \, i b_0 \theta $.

    Because variations in the phase of $t$ are equivalent to
variations in $\theta$, the $\theta$-dependence of the coefficients
$\{ c_n(\theta) \}$ is controlled by the beta function,
\begin {eqnarray}
	0 &=&
\left( 	 i {\partial \over \partial\theta} + 	 t
{\partial \over \partial t}
\right) 	 K (t e^{i\theta})
\nonumber
\\
    &\sim&
\left[ 	 - { \tilde c_{-1} \over g^2(-t) } +
\tilde c_0
\right]_{} 	{\beta (g^2(-t)) \over g^2(-t) }
\nonumber
\\
    &&
\quad {} - \tilde c_{-1} b_0 + 	\sum_{n=1}^\infty 	\left[
i {\partial \over \partial\theta} +
 n {\beta (g^2(-t)) \over
g^2(-t)}
\right] 	c_n (\theta) \> g^2(-t)^n \,.
\end {eqnarray}
Hence the coefficients $\{ c_n(\theta) \}$ satisfy the recursion
relation
\begin {equation}
    i {d \over d\theta} c_{n+1}(\theta) = - b_{n+1} \, \tilde c_{-1} +
b_n \, \tilde c_0 + \sum_{m=1}^n m \, b_{n-m} \, c_m (\theta)
\,.
\label {theta-dep}
\end {equation}
Our goal is to solve this recursion relation explicitly.  This will
enable us to relate the asymptotic behavior in the Euclidean region
where $-t$ is real and positive, to the corresponding behavior in the
Lorentzian domain where $t = s + i0^+$, with $s$ positive.  Taking the
imaginary part of the expansion (\ref {K-exp-3}) when $\theta = -\pi$,
and comparing with the expansion of the discontinuity,
\begin {equation}
    {\rm Im} \, K(s + i0^+) \sim \sum_{n=0}^\infty \, a_n \, g(s)^{2n}
\,,
\end {equation}
shows that the absorptive coefficients $\{ a_n \}$ are given by
\begin {eqnarray}
    a_n &=& {\rm Im} \, c_n (-\pi)
\\
\noalign {\hbox {for $n \geq 1$, while}}
    a_0 &=& -\pi b_0 \, \tilde c_{-1} \,.
\end {eqnarray}

\section {Borel sums and coherent states}
\label {coherent-states}

    To illustrate our method in simple terms, we first consider the
solution to the recursion relation (\ref {theta-dep}) for the special
case of a beta function which contains only one term, $ \beta (g^2) =
-b_0 g^4$.  The recursion relation (\ref{theta-dep}) simplifies to
\begin {equation}
    i {d \over d\theta} \, c_{n+1}(\theta) = n b_0 \, c_n(\theta)
\label {one}
\end {equation}
for $n \geq 1$, plus
\begin {equation}
    i {d \over d\theta} \, c_1(\theta) = b_0 \, \tilde c_0 \,.
\label {two}
\end {equation}
To solve these equations, it is convenient to regard the coefficients
as defining an abstract vector which may be represented as a state of
a simple harmonic oscillator,
\begin {equation}
    | C_\theta \rangle \equiv |0\rangle \, \tilde c_0 +
\sum_{n=1}^\infty
|n\rangle {n \over \sqrt{n!} } \, c_n(\theta)
\,.
\label {abstract}
\end {equation}
Here $|0\rangle$ is the usual ground state defined by
\begin {equation}
    a |0\rangle \equiv 0 \,,
\end {equation}
and the basis states
\begin {equation}
    | n \rangle \equiv { (a^\dagger)^n \over \sqrt{n!} } |0\rangle
\end {equation}
are eigenstates of the number operator $N = a^\dagger a$, where
$a^\dagger$ and $a$ are standard creation and annihilation operators
obeying
\begin {equation}
    \Bigl[ a , a^\dagger \Bigr] = 1 \,.
\end {equation}
Since
\begin {equation}
    a^\dagger |n\rangle = |n+1\rangle \sqrt{n+1} \,,
\end {equation}
the recursion relations (\ref {one}) and (\ref {two}) are equivalent
to a simple operator equation,
\begin {equation}
    i {d \over d\theta} |C_\theta\rangle = b_0 \, a^\dagger
|C_\theta\rangle \,,
\label {oneeq}
\end {equation}
which has the immediate solution
\begin {equation}
    | C_\theta\rangle = e^{- i b_0 \theta a^\dagger } |C\rangle \,,
\label {slt:one}
\end {equation}
where $|C\rangle$ is the initial vector with $\theta = 0$.

To express this solution in an explicit form, we utilize coherent
states defined by
\begin {equation}
    \langle z| \equiv \langle 0| e^{az} = \sum_{n=0}^\infty {z^n \over
\sqrt{n!} } \langle n| \,,
\label {cohen}
\end {equation}
which are (left) eigenvectors of the creation operator,
\begin {equation}
    \langle z | a^\dagger = z \langle z | \,. \qquad
\end {equation}
Thus the coherent state representative, defined as
\begin {eqnarray}
    C_\theta(z) &=& \langle z | C_\theta \rangle \,,
\\
\noalign {\hbox {obeys}}
    C_\theta(z) &=& e^{-i b_0 \theta z} C(z) \,.
\label {mesmart}
\end {eqnarray}
This coherent state representation,
\begin {equation}
    C_\theta(z) = \tilde c_0 +
\sum_{n=1}^\infty {n \, c_n(\theta)
\over n!} \, z^n \,,
\end {equation}
is precisely the Borel transform of the perturbative coefficients
$\{\tilde c_0 ,\, n \, c_n (\theta) \}$.  For $\theta=0$ this is the
same transform of the dispersive coefficients introduced earlier in
Eq.~(\ref {1-term C(z)}).

The absorptive coefficients may also be assembled to form an abstract
vector,
\begin {equation}
    |A\rangle = \sum_{n=1}^\infty |n\rangle {n \over \sqrt{n!} } \,
a_n \,,
\end {equation}
whose coherent state projection,
\begin {equation}
    A(z) = \langle z | A \rangle \,,
\end {equation}
gives the Borel transform (\ref {1-term A(z)}) of the absorptive
coefficients $\{ n \, a_n \}$,
\begin {equation}
    A(z) = \sum_{n=1}^\infty {n \, a_n \over n! } \, z^n \,.
\end {equation}
Using Eq.~(\ref{mesmart}) to rotate from $\theta = 0$ to $\theta =
-\pi$ (from the negative real $t$-axis back to the positive axis) and
taking the imaginary part gives
\begin {equation}
    A(z) = \sin(\pi b_0 z) \, C(z) \,.
\end {equation}
This is the result of \cite{Brown&Yaffe} quoted in Eq.~(\ref {A=sin
C}).  We have obtained it in a very simple fashion which demonstrates
an interesting and useful connection: The Borel transform is the
coherent state representation of the abstract vector which describes
the original perturbative series.

\section {General Beta Function}
\label {two-term-results}

    As was discussed earlier, the first two terms of the perturbative
expansion of the beta function are scheme independent and uniquely
determined.  On the other hand, the higher order terms are scheme
dependent.  They may be altered by using different renormalization
schemes corresponding to coupling redefinitions of the form
\begin {equation}
    \bar g^2 = g^2 + d_4 \, g^4 + d_6 \, g^6 + \cdots
\label {coupling-redefinition}
\end {equation}
which leave the lowest, order $g^2$, term unchanged.  This is
straightforward to verify directly by inserting Eq.~(\ref
{coupling-redefinition}) into
\begin {eqnarray}
    \mu^2 {d \bar g^2 \over d \mu^2} &\equiv& \bar \beta (\bar g^2)
\,,
\label {beta-bar}
\\
\noalign {\hbox {and comparing with}}
    \mu^2 {d g^2 \over d \mu^2} &\equiv& \beta (g^2) \,.
\label {beta-std}
\end {eqnarray}
As we shall see, it will prove convenient to exploit this freedom and
employ a ``two-term inverse beta function'',
\begin {equation}
    {1 \over \beta (g^2)} =\, -{1 \over b_0 g^4} + {\lambda \over b_0
g^2} \,.
\label {twodown}
\end {equation}

    Before proceeding to extend the results of the preceding section
to this case of an essentially general beta function, it is worth
pausing to describe the relation of this coupling definition to that
where the beta function itself contains only two terms,
\begin {equation}
    \bar \beta (\bar g^2) = -b_0 \, \bar g^4 - b_0 \lambda \, \bar
g^6.
\end {equation}
It is easy to check from Eqs.~(\ref {beta-bar}) and (\ref {beta-std})
that
\begin {equation}
    {1 \over g^2} = {1 \over \bar g^2} + \lambda
\label {beta -> beta-bar}
\end {equation}
converts $\beta (g^2)$ into $\bar \beta (\bar g^2)$.  The relationship
of the perturbation series for the current-current correlation
function for these two choices of the beta function is, in fact, a
simple application of the mathematical techniques developed in the
preceding section.  In view of Eq.~(\ref {coupling-expansion}), with a
one-term beta function a change of scale from $\mu_1^2$ to $\mu_2^2$
induces a change in the coupling of
\begin {equation}
    {1 \over g^2 (\mu_2^2)} = {1 \over g^2 (\mu_1^2)} + b_0 \, \ln
(\mu_2^2 / \mu_1^2) \,.
\end {equation}
This is precisely the coupling redefinition (\ref {beta -> beta-bar})
if we identify $g^2 = g^2 (\mu_2^2)$, $\bar g^2 = g^2 (\mu_1^2)$, and
$\lambda = b_0 \ln (\mu_2^2 / \mu_1^2)$.  Thus, if we replace the
phase rotation $e^{i \theta}$ used in the previous section by the
scale factor $\mu_2^2 / \mu_1^2 = e^{\lambda / b_0}$, then the
previous solution (\ref {mesmart}) of the one-term renormalization
group relations (\ref {one}), (\ref {two}) implies that
\begin {equation}
    C(z) = e^{\lambda z} \bar C(z) \,,
\end {equation}
where $C(z)$ and $\bar C(z)$ are the Borel transforms, defined by
Eq.~(\ref{1-term C(z)}), for the two different schemes.  This is the
result quoted earlier in Eq.~(\ref{zhai}).  Since the relation between
these two schemes has this simple, explicit form, it suffices to work
out the consequences of the more convenient, two-term inverse beta
function.

    We turn now to solve the recursion relation (\ref{theta-dep}) for
the two-term inverse beta function (\ref{twodown}). This we shall do
by developing an operator technique which generalizes that introduced
in the previous section. Inserting the two-term inverse beta function
(\ref{twodown}) into the recursion relation (\ref{theta-dep}) gives
\begin{equation}
    i {d \over d\theta} c_{n+1} (\theta) =
b_0(\tilde c_0
- \lambda\tilde c_{-1})\lambda^n
+ \sum_{m=1}^n m b_0
\lambda^{n-m} c_m(\theta)\,.
\end{equation}
By subtracting successive equations, this can be rewritten as
\begin{equation}
    i {d \over d\theta} \{c_{n+1}(\theta)
- \lambda
c_n(\theta)\} =
n b_0 c_n(\theta)
\label{recur:inv}
\end{equation}
for $n\geq1$, while
\begin{equation}
    i {d \over d \theta} c_1(\theta) =
b_0 (\tilde c_0 -
\lambda \tilde c_{-1})\,.
\label{recur:inv0}
\end{equation}
The generalization of the previous abstract vector definition (\ref
{abstract})
\begin{equation}
    |C_\theta \rangle = |0 \rangle (\tilde c_0 - \lambda \tilde
c_{-1})
+ \sum_{n=1}^\infty | n \rangle
{n \over \sqrt{n!}} \, c_n(\theta) \,,
\label{C:def}
\end{equation}
transcribes the relations (\ref{recur:inv}) and (\ref{recur:inv0})
into an operator equation
\begin{equation}
    \left(1 - \lambda a^\dagger {1 \over N} \right )
i{d \over
d\theta} |C_\theta \rangle
= b_0 a ^\dagger |C_\theta \rangle \,,
\end{equation}
or
\begin{equation}
    i {d \over d\theta} |C_\theta \rangle
 = b_0 S |C_\theta
\rangle \,,
\label{eq:two}
\end{equation}
where
\begin{equation}
    S = \left ( 1 - \lambda a^\dagger{1 \over N} \right )^{-1}
a^\dagger \,.
\end{equation}
Using the commutation relation $[N,a^\dagger] = a^\dagger$, this
operator may be rewritten as
\begin{eqnarray}
    S &=& a^\dagger \left ( 1 - \lambda a^\dagger {1 \over N+1}\right)^{-1}
\nonumber\\
      &=& a^\dagger (N+1) {1 \over N+1 - \lambda a^\dagger}
\nonumber\\
      &=& a^\dagger + \lambda a^\dagger a^\dagger
	  {1 \over N-\lambda a^\dagger + 1} \,.
\end{eqnarray}
Just as in the work of the previous section, Eq.~(\ref{eq:two}) has
the straightforward operator solution
\begin{equation}
    |C_\theta \rangle = e^{-ib_0\theta S} |C \rangle,
\label{slt:two}
\end{equation}
where $|C \rangle$ is the initial vector at $\theta = 0$.

Continuing to work in the spirit of the previous section, we introduce
left eigenstates of the operator $S$,
\begin{equation}
    \zeta \langle \zeta | = \langle \zeta| S\,,
\label{eigen}
\end{equation}
which reduce to coherent states at $\lambda = 0$. The projection of
the abstract vector $|C_\theta\rangle$ onto these states defines a
generalization of the coherent state representation,
\begin{equation}
{\cal C}_\theta (\zeta) = \langle \zeta | C_\theta \rangle \,.
\end{equation}
In this representation, the abstract operator relation (\ref{slt:two})
becomes a concrete relation between ordinary functions,
\begin{equation}
    {\cal C}_\theta (\zeta)
= e^{-i b_0 \theta \zeta} \, 	 {\cal
C} (\zeta) \,.
\label{phase}
\end{equation}
We again assemble the absorptive coefficients into an abstract vector,
\begin {equation}
    | A \rangle = \sum_{n=1}^\infty | n \rangle
{n
\over \sqrt{n!}} \, a_n \,,
\label{A:def}
\end {equation}
which has the generalized coherent state representation
\begin{equation}
{\cal A}(\zeta) = \langle \zeta | A \rangle \,.
\end{equation}
Using Eq.~(\ref{phase}) to rotate from the negative real $t$-axis to
the positive axis and taking the discontinuity gives
\begin{equation}
   {\cal A}(\zeta)
= \sin( \pi b_0 \zeta) \, {\cal C}(\zeta) \,.
\label {sing}
\end {equation}
This is the relation (\ref{mod A=sin C}) quoted in Sec.~II. The
functions ${\cal A}(\zeta)$ and ${\cal C}(\zeta)$ will be seen to be
precisely the modified Borel transforms of the absorptive and
dispersive coefficients.

To use this result, we need the explicit form of the generalized
coherent state representation.
To derive it, we first note that the explicit construction
of the $\langle \zeta|$ states may be obtained from their coherent
state representative $\langle \zeta | z^*
\rangle$. To obtain this representative,  we multiply
Eq.~(\ref{eigen}) from the right by $(N-\lambda a^\dagger+1) |z^*
\rangle$ and use
\begin{equation}
    {d \over dz^*} | z^* \rangle = a^ \dagger | z^* \rangle \,,
\end{equation}
which follows from the definition (\ref{cohen}), to arrive at the
differential equation
\begin{equation}
    \zeta \left [ z^*{d \over dz^*} -\lambda{d \over dz^*} + 1 \right
]
 \langle \zeta| z^* \rangle 	= \left (1 + z^*{d \over
dz^*}\right ){d \over dz^*}
 \langle \zeta |z^* \rangle\,,
\label{diffeq}
\end{equation}
which is a standard confluent hypergeometric equation in the argument
$\zeta z^*$.  The solution we need is, however, easy to obtain
directly. Since the differential equation (\ref{diffeq}) is linear in
$z^*$, it is solved by a Laplace transform involving the kernel
$e^{pz^*}$ which converts the derivative $d/dz^*$ into a
multiplication by $p$ and replaces $z^*$ by the derivative $d / dp$.
This method produces a first-order differential equation, which leads
to the solution
\begin{equation}
    \langle \zeta | z^* \rangle = \lambda \zeta \int_0^1 du \,
(1-u)^{\lambda\zeta-1} e^{\zeta uz^*} \,.
\label{sol}
\end{equation}
Because $z^*=0$ is a regular-singular point of the differential
equation (\ref{diffeq}), the other linearly independent solution is
singular at the origin.  The scalar product of a coherent state $| z^*
\rangle$ with a vector of finite norm produces an entire analytic
function of $z^*$. Hence the solution (\ref{sol}) is the proper
solution to the differential equation (\ref{diffeq}) since it defines
an entire analytic function of $z^*$.  The state $| z^* = 0 \rangle$
is the ground state $| 0 \rangle$. The solution (\ref{sol}) is
normalized so that $
\langle \zeta | 0 \rangle = 1$.

Expanding Eq.~(\ref{sol}) in powers of $z^*$, using the standard
integral representation of Euler's beta function, and using
\begin{equation}
    \langle n | z^* \rangle = {z^{* \, n} \over \sqrt{n!}}
\end{equation}
gives the number state representation of $\langle \zeta |$,
\begin{eqnarray}
    \phi_n(\zeta) &\equiv& \langle \zeta| n \rangle / \sqrt{n!}
\nonumber\\
    &=& {\Gamma(1 + \lambda \zeta) \over
\Gamma(n + 1 + \lambda
\zeta)} \, \zeta^n\,.
\label{phidef}
\end{eqnarray}
Near the origin $\zeta=0$, $\phi_n(\zeta)$ has a power series
expansion which starts from $\zeta^n$, so the functions
$\{\phi_n(\zeta)\}\,(n = 0,1,2,\cdots) $ are linearly independent and
form a complete set of analytic functions in the region
$|\zeta|<1/|\lambda|$. Making use of these functions gives
\begin{equation}
{\cal A}(\zeta) = \sum_{n=0}^\infty \, \langle \zeta | n \rangle
\langle n | A \rangle = \sum_{n=1}^\infty \phi_n(\zeta) \, n a_n \,,
\end{equation}
and
\begin{equation}
{\cal C}(\zeta) = (\tilde c_0 - \lambda \tilde c_{-1} ) +
\sum_{n=1}^\infty
\phi_n(\zeta) \, n c_n \,,
\end{equation}
which are just the results (\ref{A(z)}), (\ref{C(z)}) quoted in
Sec.~II.

We turn now to investigate the relationship between the new transform
and the standard Borel transform, and to also put our previous results
in a more general setting. The power series expansion coefficients
$\{f_n\}$ of some function $f(x)$ may be used to define an abstract
vector according to
\begin{equation}
    | F \rangle = \sum_{n=0}^{\infty}
 | n \rangle {f_n
\over \sqrt{n!}} \,.
\end{equation}
The coherent state representation produces the Borel transform,
\begin{equation}
F(z) = \langle z | F \rangle = \sum_{n=0}^\infty f_n { z^n \over n!}
\,,
\end{equation}
while the new representation produces the modified Borel transform,
\begin{equation}
{\cal F}(\zeta) = \langle \zeta | F \rangle = \sum_{n=0}^\infty f_n
\, \phi_n(\zeta) \,.
\label{modF}
\end{equation}
The relation between the modified Borel transform and the Borel
transform can be derived by using the transformation function $\langle
\zeta | z^*
\rangle$ given in Eq.~(\ref{sol}). We first note that according to
Eq.~(\ref{cohen})
\begin{eqnarray}
    \left. | {d \over dz} \rangle \langle z| \right |_{z=0}
&=&
\sum_{n=0}^\infty {1 \over n!} \,
 {a^ \dagger}^n | 0 \rangle
\langle 0 | a^n
\nonumber\\
	&=& \sum_{n=0}^\infty | n \rangle \langle n |
 = 1 \,.
\label{magicI}
\end{eqnarray}
Utilizing this identity, the relation between the two transforms can
be easily obtained,
\begin{eqnarray}
    {\cal F}(\zeta) &=& \left. \langle \zeta | F \rangle =
\langle \zeta| {d \over dz} \rangle
\langle z | F \rangle
\right|_{z=0}
\nonumber\\
    &=&
\lambda \zeta \int_0^1 du \, \left.  	(1-u)^{\lambda
\zeta -1}
e^{\zeta u {d \over dz}} 	\langle z|F \rangle
\right|_{z=0}
\nonumber\\
    &=& \lambda \zeta \int_0^1 du \,
(1-u)^{\lambda \zeta-1}
F(\zeta u) \,.
\label{intrep}
\end{eqnarray}
This expression was previously quoted in Eq.~(\ref{F -> cal F}) of
Sec.  II.

To invert this relation, so as to express the ordinary Borel transform
in terms of the modified Borel transform, we note that by Cauchy's
formula any analytic function can be expressed as a superposition of
simple poles.  Hence the inverse relation to (\ref{intrep}) may be
found by studying how to invert a simple pole.  The state $|\beta
\rangle$ defined by the coherent state representation
\begin{equation}
B(z) = \langle z | \beta \rangle = 1 +
{\beta z \over (1-\beta
z)^{1+\lambda/ \beta}} \,,
\label{zbeta}
\end{equation}
produces a simple pole in the new representation. This can be proven
from the observation that
\begin{eqnarray}
    \lambda \zeta \int_0^1 du \,
(1-u)^{\lambda \zeta -1 }
\left [ 1 + {\beta \zeta u \over
(1 - \beta \zeta
u)^{1+\lambda/\beta}} \right ] &=&
1 - {\beta \zeta \over 1 -
\beta \zeta}
\int_0^1 d \left [ {(1-u)^{\lambda \zeta} 	\over
(1 - \beta \zeta u)^{\lambda/ \beta}} \right ]
\nonumber\\
    &=& {1 \over 1 - \beta \zeta}\,.
\end{eqnarray}
Hence, in view of Eq.~(\ref{intrep}),
\begin{equation}
{\cal B}(\zeta) = \langle \zeta | \beta \rangle = {1 \over 1- \beta
\zeta } \,.
\label{pole}
\end{equation}

Having found how to invert the relation (\ref{intrep}) for a simple
pole, we can now treat the general case.  In the neighborhood of the
origin where the modified Borel transform (\ref{modF}) is assumed to
define an analytic function, Cauchy's formula may be applied,
\begin{equation}
    \langle \zeta | F \rangle
= \oint {d \zeta' \over 2 \pi i} \,
{\langle \zeta' | F \rangle
 \over \zeta' - \zeta} \,,
\end{equation}
where the contour circles about the origin with $|\zeta'| > |\zeta|$.
In view of the transformation function (\ref{pole}), this Cauchy
formula may be written as\footnote{ So as to keep the notation
uncluttered, we use the symbol $| \beta = 1/\zeta' \rangle$ to denote
the state $| \beta \rangle$ defined by Eq.~(\ref{zbeta}), but
evaluated at $\beta = 1/\zeta'$.}
\begin{equation}
    \langle \zeta | F \rangle
= \oint {d\zeta' \over 2 \pi i \zeta'}
\,
 \langle \zeta | \, \beta = 1/\zeta' \, \rangle
\langle \zeta' | F \rangle \,.
\end{equation}
This implies the formal completeness relation%
\footnote
    {%
    This representation of the identity as a contour integral is a
generalization of an idea of Schwinger \cite{Schwinger}.  }
\begin{equation}
    \oint {d \zeta \over 2 \pi i \zeta} \,
 | \, \beta = 1/ \zeta
\, \rangle \langle \zeta |
= 1 \,.
\label{ident}
\end{equation}
This expression of the identity holds when it is inserted in matrix
elements and the contour chosen appropriately so as to enclose the
relevant singularities.  Inserting the completeness relation between
$\langle z |$ and $| F \rangle$ yields the inverse relation to
Eq.~(\ref{intrep}),
\begin{equation}
F(z) = \langle z | F \rangle = \oint {d \zeta' \over 2 \pi i \zeta'}
\,
 \langle z | \, \beta = 1/\zeta' \, \rangle
\langle \zeta' | F \rangle \,.
\label{INV}
\end{equation}
This result was presented in Eq.~(\ref{cal F -> F}) in Sec.~II, where
it was derived in a different fashion.

Number state matrix elements of the completeness relation
(\ref{ident}) give
\begin{equation}
    \oint {d \zeta \over 2 \pi i \zeta} \,
\chi_m(1/\zeta)
\phi_n(\zeta)
= \delta_{m,n} \,,
\label{circle}
\end{equation}
where
\begin{equation}
    \chi_n(\beta) \equiv \langle n | \beta \rangle \sqrt {n!} \,.
\end{equation}
Here the contour must encircle the origin with a radius constrained by
$| \zeta | < 1/ \lambda$ so as to avoid the singularities of
$\phi_n(\zeta)$.  With this restriction, Eq.~(\ref{circle}) describes
the way in which $\{\phi_n(\zeta)\}$ and $\{\chi_n(\beta)\}$ form
reciprocal sets of basis functions.  Expanding the coherent state
representative $\langle z | \beta
\rangle$  given by Eq.~(\ref{zbeta}) in
powers of $z$ identifies the number state components $\langle n |
\beta \rangle$ and yields the explicit form
\begin{equation}
 \chi_n(\beta) =  \delta_{n,0} + n \beta^n \,
{\Gamma(n + \lambda/ \beta) \over \Gamma(1 + \lambda/ \beta) } \,,
\label{chidef}
\end{equation}
which are polynomials in $\beta$. This is the result quoted in
Eq.~(\ref{chi_n}) of Sec.~II.

The completeness relation (\ref{ident}) can also be exploited to
extract the coefficients $\{f_n\}$ of an asymptotic series from the
modified Borel transform:
\begin{eqnarray}
    f_n = \langle n | F \rangle \sqrt{n!}
&=& \oint {d \zeta
\over 2 \pi i \zeta} \,
 \langle n | \, \beta = 1 /\zeta \,
\rangle
 \langle \zeta | F \rangle \sqrt{n!}
\nonumber\\
	&=& \oint {d \zeta \over 2 \pi i \zeta} \,
 \chi_n
(1/\zeta) \langle \zeta | F \rangle \,.
\label{stupid}
\end{eqnarray}
Since $ \langle \zeta | F \rangle = {\cal F}(\zeta) $, this is the
formula (\ref{F->f_n}) stated in Sec.~II.

With these results in hand, we may now derive the explicit relation
between the dispersive coefficients $\{c_n\}$ and the absorptive
coefficients $\{a_n\}$.  Applying (\ref{stupid}) to the absorptive
transform and using the simple relation (\ref{sing}) between their
modified Borel transforms, ${\cal A}(\zeta) = \langle \zeta | A
\rangle = \sin (\pi b_0 \zeta)
\langle \zeta | C \rangle $,
gives
\begin{eqnarray}
    \langle n| A \rangle \sqrt{n!} &=& \oint
    {d \zeta \over 2 \pi i \zeta}
    \, \chi_n (1/\zeta) \sin (\pi b_0 \zeta)
    \langle \zeta | C \rangle
\nonumber\\
       &=& \oint {d \zeta \over 2 \pi i \zeta}
\, \chi_n (1/\zeta)
    \left \{ \sin (\pi b_0 \zeta)
    \sum_{m=0}^\infty \phi_m( \zeta)
    \langle m| C \rangle \sqrt{m!} \right \} \,,
\label{An:Cn}
\end{eqnarray}
where in the second line Eq.~(\ref{phidef}) has been used. Conversely,
\begin{equation}
    \langle n| C \rangle \sqrt{n!} = \oint {d \zeta \over 2 \pi i
\zeta}
\, \chi_n (1/\zeta) \left \{{1 \over \sin (\pi b_0 \zeta)}
\sum_{m=0}^\infty \phi_m(\zeta)
\langle m | A \rangle
\sqrt{m!} \right \} \,.
\label{Cn:An}
\end{equation}
Using the definitions~(\ref{C:def}) and (\ref{A:def}) of the states $|
C \rangle$ and $| A \rangle$, and the explicit forms~(\ref{phidef})
and (\ref{chidef}) for the functions $\phi_n(\zeta)$ and
$\chi_n(\zeta)$ gives, for $n \geq 1$,
\begin{eqnarray}
    n a_n =&& \oint {d \zeta \over 2 \pi i \zeta}\,
{n\Gamma(n+
\lambda \zeta)
\over \Gamma(1+ \lambda \zeta)} \, 	\zeta^{-n}
\sin(\pi b_0 \zeta)
\left \{ (\tilde c_0 - \lambda \tilde c_{-1})
+
\sum_{m=1}^\infty \phi_m (\zeta)\, m c_m \right \}
\nonumber\\
    =&& \!  \sum_{m=0}^\infty
\oint {d \zeta \over 2 \pi i \zeta} \,
{n\Gamma(n {+} \lambda \zeta)
\over \Gamma(m {+} 1 {+} \lambda
\zeta)} \,
\zeta^{m-n} \sin(\pi b_0 \zeta) 	\left
\{(\tilde c_0 {-} \lambda \tilde c_{-1})
\delta_{m,0} + m c_m
\right \} .
\label{sum}
\end{eqnarray}
The terms in the summation in Eq.~(\ref{sum}) with $m \geq n$ vanish
since the integrand is then regular at $\zeta = 0$.  To evaluate this
expression, we define combinatorial factors $I_{m,l}^n$ (where $n, m,
l$ are nonnegative integers which satisfy $n \geq m+l$) by the
generating function
\begin{equation}
    {\Gamma(n+1+x) \over \Gamma(m+1+x)}
= \sum_{l=0}^{n-m}
x^{n-m-l} I_{m,l}^n\,.
\label{clever}
\end{equation}
Inserting this expansion in Eq.~(\ref{sum}) and expanding $\sin(\pi
b_0 \zeta)$ in powers of $\zeta$ now yields
\begin{eqnarray}
    a_n = &&
\sum_{m=0}^{n-1} 	\oint {d \zeta \over 2 \pi i
\zeta }
\left ( \sum_{k=0}^\infty (-)^k 	{(\pi b_0
\zeta)^{2k+1} \over (2k{+}1)! } \right )
\nonumber\\
	&& \qquad \times \sum_{l=0}^{n-m-1}
(\lambda
\zeta)^{n-m-l-1} \zeta^{m-n} I_{m,l}^{n-1}
\left \{ (\tilde c_0 -
\lambda \tilde c_{-1})
\delta_{m,0} + m c_m \right \}
\nonumber\\
    =&&
(\tilde c_0 - \lambda \tilde c_{-1}) 	\left
\{\sum_{k=0}^{[{n-1 \over 2}]}
(-)^k {(\pi b_0)^ {2k{+}1} \over
(2k+1)!}\,
\lambda^{n-2k-1} I_{0,2k}^{n-1} \right \}
\nonumber\\
    && \quad +
\sum_{m=1}^{n-1} m c_m \left \{
\sum_{k=0}^{[{n-m-1 \over 2}]}
(-)^k {(\pi b_0)^{2k{+}1} \over
(2k+1)!} \,
\lambda^{n-m-2k-1} I_{m,2k}^{n-1} \right \} \,.
\end{eqnarray}
Here $[x]$ denotes the integer part of $x$.  This result provides an
explicit evaluation of the absorptive coefficients $a_n$ in terms of
the dispersive coefficients $\{c_m\}$ with smaller indices, $m < n$.
It is the result (\ref{a_n = sum c_n}) of Sec.~II. In the limit
$\lambda \to 0$, using
\begin{equation}
    I_{n-k,k}^n = {\Gamma(n+1) \over \Gamma(n-k+1)}
= {n! \over
(n-k)!} \,,
\end{equation}
we find that for $n > 0$
\begin{equation}
    {a_n \over (n{-}1)!}
= \delta_{n,{\rm odd}} \, (-)^{{n-1
\over 2}}
 {(\pi b_0)^n \over n!} \, \tilde c_0 	+
\sum_{k=0}^{[{n \over 2}]-1}
 (-)^k {(\pi b_0)^{2k+1} \over
(2k{+}1)!} \,
 {c_{n-2k-1} \over (n{-}2k{-}2)!} \,,
\end{equation}
which is precisely the previous result in \cite{Brown&Yaffe} as quoted
in Eqs.~(\ref{a_even}) and (\ref{a_odd}) of Sec.~II.

Making use of the expansion
\begin{equation}
    {1 \over \sin\,z}
= \sum_{k=0}^\infty 	{|(2^{2k} -2) B_{2k}|
\over (2k)!}\, z^{2k-1} \,,
\end{equation}
where $B_n$ are the Bernoulli numbers, and going through similar
steps, one may expresses the dispersive coefficients $\{c_n\}$ as a
sum of the absorptive coefficients $\{a_n\}$.  For $n > 1$, one has
\begin{eqnarray}
    c_n &=&
\sum_{m=1}^{n+1} m a_m 	\left \{ \oint {d \zeta \over
2 \pi i \zeta} \,
{\Gamma(n {+} \lambda \zeta) 	\over \Gamma(m
{+} 1 {+} \lambda \zeta)} \,
{1 \over \sin \pi b_0 \zeta} \,
\zeta^{m-n} \right \}
\nonumber\\
	&=& { (a_{n+1} {-} \lambda a_n)
 \over n \pi b_0}
+ \sum_{m=1}^{n-1} m a_m
 \oint {d \zeta \over 2 \pi i \zeta} {
1 \over \sin \pi b_0 \zeta } \sum_{l=0}^{n-m-1}
 I_{m,l}^{n-1}
\, (\lambda \zeta)^{n-m-l-1}
 \zeta^{m-n}
\nonumber\\
	&=& {(a_{n+1} {-} \lambda a_n )
\over \pi b_0 \, n} +
\sum_{m=1}^{n-1} m a_m
\sum_{k=1}^{[{{n-m+1} \over 2}]}
{|(2^{2k} - 2) B_{2k}| \over (2k)!}
(\pi b_0)^{2k-1}
\lambda^{n-m-2k+1} I_{m,2k-2}^{n-1} \,,
\nonumber\\
\label{done}
\end{eqnarray}
while
\begin{equation}
    \tilde c_{-1} = -{a_0 \over \pi b_0} \,, \qquad \tilde c_0 = {a_1
- \lambda a_0 \over \pi b_0} \,.
\label{finished}
\end{equation}
This is the result displayed in Eq.~(\ref{c_n = sum a_n}) of Sec.~II.
Again the dispersive coefficient $c_n$ involves only the absorptive
coefficients $\{a_m\}$ with $m \leq n+1$ and the $\lambda \to 0$ limit
reproduces the previous result in \cite{Brown&Yaffe} quoted in
Eq.~(\ref{nc_n}).

\acknowledgments

We should like to thank G. West for initially sparking our interest in
this subject.  One of the authors (L. S. B.) would like to acknowledge
the hospitality of the Los Alamos National Laboratory where part of
his research for this paper was performed.  The work was supported, in
part, by the U. S. Department of Energy under grant DE-AS06-88ER40423.

\newpage

\appendix{critique of West's paper}

    West \cite{West} considers the differentiated
correlation function $t \, d K(t) / dt$ which he terms
$D(t/\mu^2, g^2(\mu^2))$ [defined in his Eq.~(8)].%
\footnote
    {%
    Some of the properties of this function are
    discussed in Sec.~III, Eqs.~(\ref{deriv})--(\ref{mayday}).
    }
This is a renormalization group invariant
 which has a perturbative expansion in powers of the running
coupling,
\begin {equation}
    D (t/\mu^2, g^2(\mu^2)) \sim
	\sum_{n=0}^{\infty} \, (-)^n d_n(t/\mu^2) \,g(\mu^2)^{2n}  \,.
\label {asymp}
\end {equation}
[To facilitate comparison with West's paper, we use his notation
for the coefficients $\{ d_n \}$ which, for $\mu^2 = - t$,
differ from our $\{ k'_n \}$ in Eq.~(\ref {deriv}) by a factor $(-)^n$.]
West asserts (in his Eq.~(20))
that these coefficients have the large order behavior
\begin {equation}
    d_n(t/\mu^2) \sim \left[{2 \over \pi \phi(k_1)}
		      \right]^{1/2} k_1^{n-1}
		      D(t/\mu^2, 1/k_1) \cos n \pi \,,
\label {wrong}
\end {equation}
where, in our notation, $k_1 \sim b_0(n{-}1) + \lambda$,
and $\phi(k_1) \sim (b_0 k_1)^{-1}$.
By using the free field limit
$D(t/\mu^2, 0) =  -\sum_f Q_f^2/ 4\pi^2$,
where $\{ Q_f \}$ are the quark charges
(which follows from Eq.~(8) in \cite{West}),
the claimed large order behavior of the coefficients $d_n(t/\mu^2)$
can be written explicitly as
\begin {equation}
    d_n(t/\mu^2) \sim - { \sqrt{2\pi} \over 4\pi^3} e^{-1+\lambda/b_0}
		 \Bigl( \sum\nolimits_f Q_f^2 \Bigr)
		 (-b_0)^n\, n^{n-1/2}\,.
\label {ExpWest}
\end {equation}
We shall show that this result is both inconsistent and contradicts
our exact relations.

    Because $D(t/\mu^2,g^2(\mu^2))$ satisfies the
homogeneous renormalization group equation
\begin {equation}
    \mu^2 {d \over d \mu^2} D(t/\mu^2,g^2(\mu^2)) = 0\,,
\label {homoRG}
\end {equation}
the coefficients in the expansion (\ref{asymp}) must obey
\begin {equation}
    \sum_{n=0}^{\infty}\, (-)^n g^{2n}(\mu^2) \,
		\left \{ \mu^2 {\partial \over \partial \mu^2}
		+ {n \beta(g^2(\mu^2))
		 \over g^2(\mu^2)} \right \}  d_n(t/\mu^2) = 0\,.
\label {Diff}
\end {equation}
Using the asymptotic expansion of the beta function,
$\beta(g^2) \sim -b_0 \, g^4 - b_0 \lambda \, g^6 - \cdots \,,$
and identifying the coefficients of each power of the coupling yields
\begin {equation}
    \mu^2 {\partial \over \partial \mu^2} d_n(t/\mu^2)
	= -b_0 \left \{ (n{-}1) d_{n-1} (t/\mu^2)
	- \lambda (n{-}2) d_{n-2} (t/\mu^2) + \cdots \right \} \,.
\label {RR}
\end {equation}
Using the explicit large order behavior (\ref{ExpWest}) on the
right-hand side of Eq.~(\ref{RR}) implies that
\begin {equation}
    \mu^2 {\partial \over \partial \mu^2} d_n(t/\mu^2)
	\sim  {\rm const.}\, (-b_0)^n n^{n-1/2} \,,
\end {equation}
which states that the logarithmic derivative $(\mu^2 d/d \mu^2)d_n(t/\mu^2)$
has the same large order behavior as does $d_n(t/\mu^2)$ itself.
However, this contradicts Eq.~(\ref{ExpWest}) which asserts that
the leading order behavior of $d_n(t/\mu^2)$ is independent of $\mu^2$.

    We shall now show that formula (\ref{wrong}) also contradicts our results.
The assertion (\ref{ExpWest}) may be recast as a prediction for the dispersive
coefficients $\{c_n\}$. Comparing the integrated expansion of $t\,dK/dt$,
Eq.~(\ref{away}), to the original Euclidean expansion (\ref{K-exp-2}) and
using our two-term inverse beta function~(\ref{recycle}) yields
\begin{eqnarray}
   n c_n &=& - (k_{n+1}' - \lambda k_n')/b_0
\nonumber\\
         &=& (-)^n \{ d_{n+1}(-1) + \lambda d_n(-1) \}/b_0 \,.
\label{Dislarg}
\end{eqnarray}
This implies that the perturbative coefficients
$\{k_n'\}$ of the differentiated correlation function have
the same large-order behavior as that of the original coefficients
$\{c_n\}$ (up to an overall constant factor), provided that the
coefficients have $n!$ growth. If the differentiated coefficients
have the asymptotic behavior (\ref{ExpWest}), then from
Eq.~(\ref{Dislarg}), the dispersive coefficients will satisfy
\begin{eqnarray}
    c_n &\sim& {(-)^n \over n b_0} \, d_{n+1}(-1) \times (1 + O(1/n))
\nonumber\\
	&\sim& {\sqrt{2 \pi} \over 4 \pi^3} \,
e^{\lambda/b_0} \Bigl( \sum\nolimits_f Q_f^2 \Bigr) \, b_0^n  \, n^{n-1/2}
\nonumber\\
	&\sim& {e^{\lambda /b_0} \over 4 \pi^3}
	\Bigl( \sum\nolimits_f Q_f^2 \Bigr) (e b_0)^n \, \Gamma(n) \,.
\label {largdis}
\end{eqnarray}

Such large order behavior (\ref{largdis}) for the dispersive coefficients
creates a singularity in the Borel transform
\begin{equation}
    C(z) = \tilde c_0 - \lambda \tilde c_{-1} +
	 \sum_{n=1}^{\infty} {n c_n \over n!} \, z^n
\label{BorelC}
\end{equation}
at $z = 1/e b_0$. (No such singularity, closer to the origin than
the first ultraviolet renormalon at $b_0 z = -1$, has been found
in any investigation of individual Feynman diagrams
\cite{Mueller}.)
Inserting the asymptotic form (\ref{largdis}) into Eq.~(\ref{BorelC})
yields the leading behavior of $C(z)$ near $z = 1/eb_0$,
\begin{equation}
    C(z) \sim {e^{\lambda/b_0} \over 4 \pi^3}
	 \Bigl( \sum\nolimits_f Q_f^2 \Bigr) \, (1 - e b_0 z)^{-1} \,.
\label{Csingul}
\end{equation}

We now go through similar steps for the absorptive coefficients ${a_n}$.
Since Eq.~(\ref{Dislarg}) is the special case of the general relation
between the phase-dependent coefficients $\{ c_n(\theta)\}$ and the
momentum-dependent coefficients $\{ d_n(t/|t|)\}$, with $t = -|t|
e^{ i \theta }$, the analytic continuation of Eq.~(\ref{Dislarg})
produces
\begin{equation}
   n a_n = (-)^n  \left\{ {\rm Im}\, d_{n+1} (1)
	  + \lambda \, {\rm Im}\, d_n(1) \right\} /b_0 \, .
\end{equation}
Equation~(21) in \cite{West} gives the predicted
large order behavior\footnote{
	This differs from the original form of Eq.~(21) in \cite{West}
by a factor of $3 (\sum\nolimits_f Q_f^2)$, which we believe was missing
there.} of ${\rm Im}\, d_n(1)$,
\begin{eqnarray}
    {\rm Im}\, d_n(1) &\sim& (-)^{n+1} { b_0 \over 8 \pi^3}
	\Bigl( \sum\nolimits_f Q_f^2 \Bigr) \,
	{ k_1^{n-3} \over \sqrt{2 \pi \phi(k_1)}}
\nonumber\\
	&\sim&  (-)^{n+1} { e^{-1 + \lambda /b_0}
	 \over 8 \pi^3 \sqrt{2 \pi}}
	\Bigl( \sum\nolimits_f Q_f^2 \Bigr) \, b_0^{n-1} n^{n-5/2} \,.
\end{eqnarray}
Hence the large order behavior of the absorptive coefficients ${a_n}$
is given by
\begin{eqnarray}
    a_n &\sim& {e^{\lambda/b_0} \over 8 \pi^3 \sqrt{2 \pi}}
	\, \Bigl( \sum\nolimits_f Q_f^2 \Bigr) \, b_0^{n-1}\, n^ {n-5/2} \,,
\nonumber\\
&\sim& {e^{\lambda/b_0} \over 16 \pi^4 b_0}
	\, \Bigl( \sum\nolimits_f Q_f^2 \Bigr) \, (e b_0)^n \, \Gamma (n{-}2) \,.
\label{largabs}
\end{eqnarray}
Comparing the large order behavior of ${c_n}$ given in Eq.~(\ref{largdis})
with that of ${a_n}$ given in Eq.~(\ref{largabs}),
we find that $a_n/c_n \sim 1/n^2$, which
contradicts the constraint, described in Sec.~IIE, that $a_n$
and $c_n$ must have the same large-order behavior if the radius of
convergence of the Borel transform is less than $1/b_0$.
Hence (\ref {wrong}) cannot be correct.

Incidentally, we also find that substituting Eq.~(21) into Eq.~(12) in
\cite{West}
does not give the large order behavior for $r_n(1)$ stated in Eq.~(22).
Using $\alpha_s = g^2(s)/ 4 \pi$ and
\begin{equation}
    R(s) \sim 3 \Bigl( \sum\nolimits_f Q_f^2 \Bigr) \,
	 \sum_{n=0}^{\infty} \,
	 r_n(1) \left( {\alpha_s(s) \over \pi}\right)^n \,,
\end{equation}
we find from Eq.~(\ref{largabs}) that
\begin{equation}
    r_n(1) \sim {e^{\lambda/b_0} \over 4\pi^3 b_0}
	(4 \pi^2 e b_0 )^n \, \Gamma (n{-}2) \,,
\label{corrected}
\end{equation}
which differs from Eq.~(22) in \cite{West} by the factor
$-n^{\lambda/b_0}$.
However, the corrected result (\ref{corrected}) must still be
in error as it arises from the inconsistent result (\ref{wrong}).

\appendix {Borel transform details} \label {Borel asymptotics}

We show first the equivalence of the singularities of the modified Borel
transform and the standard Borel transform presented in
Eqs.~(\ref{mod F sing}) and (\ref{mod F sing -> F sing}).
Suppose that the Borel transform $F(z)$ has a singularity at
$z = R$ of the form
\begin{equation}
    F(z) = (1 - z/R)^{-\alpha-\lambda R}\,.
\label{Assum}
\end{equation}
Inserting Eq.~(\ref{Assum}) into Eq.~(\ref{F -> cal F}) gives
the modified Borel transform
\begin{equation}
    {\cal F} (\zeta) = \lambda \zeta \int_0^1 du \,
			(1-u)^{\lambda \zeta-1}
			(1 - \zeta u/R) ^{-\alpha-\lambda R} \,,
\label{standard-rep}
\end{equation}
which is Euler's integral representation of the hypergeometric
function \cite{Bateman},
\begin{equation}
    {\cal F} (\zeta) = F (\alpha + \lambda R,1;\lambda \zeta +1;
			\zeta / R) \,.
\label{hygeom}
\end{equation}
Since the hypergeometric function $F(a,b;c;z)$ is analytic in the
domain where $|z| < 1$, to examine the behavior of ${\cal F}(\zeta)$
near $\zeta = R$, we make use of the analytic continuation
\begin{eqnarray}
    F(a,b;c;z) =&& A_1\,  F(a, b; a+b-c+1; 1-z)
\nonumber\\
		&&{}+ A_2 \, (1-z)^{c-a-b} F(c-a,c-b;c-a-b+1;1-z) \,,
\end{eqnarray}
where
\begin{equation}
    A_1 = {\Gamma(c) \Gamma(c-a-b)
	\over \Gamma(c-a) \Gamma(c-b)} \,,
	\qquad A_2 = {\Gamma(c) \Gamma(a+b-c)
	\over \Gamma(a) \Gamma(b)} \,.
\end{equation}
Noting that
\begin{equation}
F(a, b; a; z) = (1 - z)^{-b} \,,
\end{equation}
we now find that the modified Borel transform ${\cal F}(\zeta)$ may be
expressed as
\begin{equation}
    {\cal F} (\zeta) = {\lambda \zeta \over \eta} \,
	F( \lambda R + \alpha, 1; 1 - \eta; 1 - \zeta/R)
	+ {\Gamma(\lambda \zeta + 1)
	\Gamma(-\eta) \over \Gamma( \lambda R + \alpha)} \,
	(1 - \zeta/R)^{\eta} (\zeta/R)^{-\lambda \zeta} \,,
\label{analytic}
\end{equation}
where $\eta = \lambda \zeta {-} \lambda R {-} \alpha$.
Since $F(a,b;c;0) = 1$,
Eq.~(\ref{analytic}) gives the leading singular behavior of
${\cal F}(\zeta)$ as $\zeta \to R$ as well as the first correction,
\begin{eqnarray}
    {\cal F}(\zeta) &\sim& {\Gamma(\lambda R +1) \Gamma(\alpha)
			    \over \Gamma (\lambda R + \alpha)} \,
			    (1{-}\zeta / R)^{-\alpha}
\nonumber\\
	&& \qquad {}\times \Bigl[ 1 - \lambda R \, (1{-}\zeta/R)
			    \ln (1{-}\zeta/R)
			+  O (1 {-} \zeta/R) \Bigr] \,.
\label{log}
\end{eqnarray}

This shows that if the Borel transform $F(z)$ is exactly a
simple power law singularity, then the
modified Borel transform ${\cal F}(\zeta)$ will contain a
singularity at the same position but with a shifted power, and with
subleading corrections suppressed by $ (1{-}\zeta/R) \ln (1{-}\zeta/R)$. It is
not difficult to see that the converse also holds: If the modified
Borel transform ${\cal F}(\zeta)$ has only the first power-law term
shown in Eq.~(\ref{log}), then the Borel transform $F(z)$ will have
the singularity shown in Eq.~(\ref{Assum}) with a subleading correction
suppressed by $ (1{-}z/R) \ln (1{-}z/R)$.
The equivalence we have demonstrated is just that between
Eqs.~(\ref{mod F sing}) and (\ref{mod F sing -> F sing}) stated in
the text.

We now turn to prove that the
transformation~(\ref{inverse-mod-Borel-a}) constructs from the
modified Borel transform ${\cal F} (z)$ the same ``inverse Borel
transform'' $f(y)$ as is produced by the
Laplace transform~(\ref{inverse-Borel-a})
of the standard Borel transform.
Substituting the expression~(\ref{cal F -> F}) for the Borel
transform $F(z)$ in terms of the modified Borel
transform ${\cal F} (z)$ into the Laplace
transform~(\ref{inverse-Borel-a}) gives
\begin{equation}
    f(y) = {1 \over y} \int_0^{\infty} dz \,e^{-z/y}
	   \oint {d \zeta \over 2 \pi i \zeta}
	   \left ( 1 + {z/\zeta \over
	   (1-z/\zeta)^{1+\lambda \zeta}}
	   \right ) {\cal F} (\zeta) \,.
\end{equation}
Interchanging the order of the
integrals\footnote{
This is valid under the condition that the integrals
converge absolutely. For $\lambda \leq 0$ it is sufficient
that the modified Borel transform ${\cal F}(\zeta)$
be bounded and analytic within a neighborhood of the contour
of the $z$-integral (assumed to lie in the right half plane).
For $\lambda > 0$ sufficient conditions are that ${\cal F}
(\zeta)$ be bounded and analytic within some wedge enclosing
the contour of the $z$-integral (so that $z/\zeta$ may remain
bounded away from one as both $z$ and $\zeta \to \infty$).
For sufficiently small values of $y$ the integrand is then
exponentially bounded.
    }
yields
\begin{equation}
     f(y) = {1 \over y} \oint_C
	    {d \zeta \over 2 \pi i \zeta}
	    \, {\cal F} (\zeta) \int_0^{\infty}
	    \, dz\, e^{-z/y}
	    \left( 1 + {z/\zeta \over
	    (1 - z/\zeta)^{\lambda \zeta +1}}
	    \right) \,,
\end{equation}
where the contour $C$ wraps counterclockwise about the entire
path of the $z$-integral.
Separating the integrand into two pieces by writing
\begin{equation}
   1 +  {z/\zeta \over (1-z/\zeta)^{1+\lambda \zeta}}
	= {1 \over (1-z/\zeta)^{1+\lambda \zeta}} +
	\left\{ 1 - {1 \over (1-z/\zeta)^{\lambda \zeta}} \right\} \,,
\end{equation}
and then, for the term in braces, integrating by parts in $z$ yields
\begin{equation}
    f(y) = {1 - \lambda y \over y}
	\oint_C {d \zeta \over 2 \pi i \zeta}
	\, {\cal F} (\zeta) \int_0^{\infty} dz \, e^{-z/y}
	    \left \{ {1 \over (1 - z / \zeta)^{1+\lambda \zeta}}
	    \right \} \,.
\label{Hidden}
\end{equation}
The $\zeta$-contour integral can be written as a line integral of
${\cal F}(\zeta)$ times the discontinuity (in $\zeta$) of the
function $\int_0^{\infty} dz \, e^{-z/y}
(1 - z/\zeta)^{-1-\lambda \zeta}$. Hence
\begin{equation}
    f(y) = - {1 - \lambda y \over y} \int_0^{\infty} d \zeta
	\,{\cal F} (\zeta) \oint_C {dz \over 2 \pi i \zeta} \,
	e^{-z/y} (1 - z/\zeta)^{-1 - \lambda \zeta}\,,
\end{equation}
where the path of the $\zeta$-integral is now the same as the
original path of $z$. The change of variable $z = yt + \zeta$
converts the inner integral into the standard representation
of an inverse gamma function,
\begin{eqnarray}
    f(y) &=& {1 - \lambda y \over y} \int_0^{\infty}
		d \zeta \, e^{- \zeta/y}
		(\zeta/y)^{\lambda \zeta}\,{\cal F} (\zeta)
		\,{i \over 2\pi} \int_C dt \, e^{-t} \,
		(-t)^{- \lambda \zeta -1}
\nonumber\\
	&=& {1 - \lambda y \over y} \int_0^{\infty}
		d \zeta \, e^{- \zeta/y}
		{ (\zeta/y)^{\lambda \zeta}
		\over \Gamma( 1 + \lambda \zeta)}
		\, {\cal F} (\zeta)\,,
\label{Wonderful}
\end{eqnarray}
which is the result (\ref{inverse-mod-Borel-a}).

We can also show directly that the transformation~(\ref{Wonderful})
defines a function which has the desired expansion.
After rescaling $\zeta$ by a factor of $y$,
the representation~(\ref{Wonderful})
immediately shows that $f(y)$ has an asymptotic expansion in powers
of $y$, $f(y) \sim
\sum_{n=0}^{\infty} c_n \, y^n$. The coefficients may be computed by
inserting the definition~(\ref{Yaffe}) of ${\cal F}$
into Eq.~(\ref{Wonderful}), substituting $\zeta = sy$,
\begin{equation}
    f(y) = (1{-}\lambda y) \int_0^{\infty} ds \,
	   e^{-s}\,s^{\lambda sy} \sum_{m=0}^{\infty}\,
	   f_m \,{ (sy)^m \over \Gamma (m{+}1{+}\lambda sy)}\,,
\end{equation}
and noting that only the first $n$ terms in the sum can
contribute terms of order $y^n$.
Hence the coefficient $c_n$ may be extracted by a contour integral
encircling the origin,
\begin{eqnarray}
    c_n &=& \oint {dy \over 2 \pi i y}
	    {1{-}\lambda y \over y^n}
	    \int_0^{\infty} ds \, e^{-s} \,
	    s^{\lambda s y} \sum_{m=0}^n\,
	   f_m \,  { (sy)^m \over \Gamma(m{+}1{+}\lambda s y)}
\nonumber\\
	&=& \sum_{m=0}^n f_m \oint {dz \over 2 \pi i z}
	    {(z/\lambda)^{m-n} \over \Gamma(m{+}1{+}z)}
	    \int_0^{\infty} ds\,e^{-s}\,s^{z+n} (1{-}z/s)
\nonumber\\
	&=& \sum_{m=0}^n f_m \oint
	    {dz \over 2 \pi i z} (z/\lambda)^{m-n}
	    {n \,\Gamma(n{+}z) \over \Gamma(1{+}m{+}z)} \,.
\end{eqnarray}
The terms with $m<n$ give no contribution since, for these terms,
the ratio of gamma functions is a polynomial in $z$ of order
$n{-}m{-}1$ and thus when divided by $z^{n-m+1}$ produces a vanishing
residue at the origin. The final term gives
\begin{equation}
    c_n = f_n \oint {dz \over 2 \pi i z}
	  {n \over z+n} = f_n \,,
\end{equation}
so that, as expected, $f(y) \sim \sum_{n=0}^{\infty} f_n \, y^n$.

\appendix {Alternative derivations} \label {Alternate derivations}

    Several different approaches may be used to derive the relations
between the absorptive and dispersive perturbative expansion coefficients.
This appendix sketches two complementary methods which do not
involve the abstract vector representations employed in the main text.

    By examining the asymptotic behavior of the dispersion relation
satisfied by the correlation function $K(t)$, one may derive
the result (\ref {c_n = sum a_n}) expressing the dispersive coefficients
in terms of the absorptive coefficients.
The scalar correlation function $K(t)$ satisfies the once-subtracted
dispersion relation
\begin {equation}
    \Delta K (t) \equiv K(t) - K (0) = {t \over \pi} \,
    \int_0^\infty {ds \over s} \, {\rho (s) \over (s-t) } \,,
\label {dispersion-relation}
\end {equation}
where the spectral density $\rho(s)$ is the discontinuity of
$K(t)$ across the positive real axis,
\begin {equation}
    \rho (s) \equiv {\rm Im} K(s+i0^+) \,.
\end {equation}
Given the asymptotic expansion of the spectral density,
\begin {equation}
    \rho (s) \sim \sum_{n=0}^\infty \> a_n \, (g^2(s))^n \,,
\label {a_n exp}
\end {equation}
this dispersion relation may be used to derive the
asymptotic behavior of the correlation function $K(t)$ as $|t| \to \infty$
along any ray in the complex $t$-plane.
To carry this out, it is convenient to separate the dispersion integral
(\ref {dispersion-relation})
into two pieces, $0 < s < |t|$, and $|t| < s < \infty$.
In the low momentum piece, one may expand the integrand in powers of $s/t$,
and for the high momentum piece, expand in powers of $t/s$.
Hence
\begin {eqnarray}
    \pi \, \Delta K (t)
    &=& - \int_0^{|t|}      {ds \over s} \, \rho (s)
	    \left( \sum_{n=0}^\infty (s/t)^n \right)
	+ \int_{|t|}^\infty {ds \over s} \, \rho (s)
	    \left( \sum_{n=1}^\infty (t/s)^n \right)
\nonumber
\\
    &\equiv& -I^-_0(t) + \sum_{n=1}^\infty \> (I_n^+(t) - I_n^-(t)) \,,
\end {eqnarray}
where
\begin {eqnarray}
    I_n^-(t) &\equiv& \int_0^{|t|} {ds \over s} \, \rho (s) \, (s/t)^n \,,
\\
\noalign {\hbox {and}}
    I_n^+(t) &\equiv& \int_{|t|}^\infty {ds \over s} \, \rho (s) \, (t/s)^n \,.
\end {eqnarray}

    The asymptotic expansion of $I_n^\pm(t)$ may be computed by
inserting the expansion (\ref {a_n exp}) of the spectral density,
changing variables from $s$ to $g^2(s)$,
and performing the resulting integrals term-by-term.
The final expressions are simplified if
one uses the rescaled inverse coupling
\begin {equation}
    z \equiv 1 \bigm/ (b_0 \, g^2(s))
\end {equation}
as the variable rather than $g^2(s)$ itself.
The relation (\ref {qsquare})
between $s$ and $g^2(s)$ may then be expressed as
\begin {equation}
    s = \Lambda^2 \, z^\gamma e^z  \,,
\end {equation}
where we have defined $\gamma \equiv -\lambda / b_0 = -b_1 / b_0^2$.
Similarly, let $y \equiv 1 / (b_0 \, g^2(|t|))$, and $t = |t| e^{i\phi}$.
Using $ds / dz = s (1 + \gamma / z)$, one finds for the high-momentum
contribution
\begin {eqnarray}
    I_n^+(t)
    &\sim& \displaystyle
	\int_y^\infty dz \> (1 + \gamma / z) \,
	\left( \sum_{m=0}^\infty a_m \, (b_0 z)^{-m} \right)
	(t/ \Lambda^2)^n \,
	z^{-n \gamma} \, e^{-n z}
\nonumber
\\
    &\sim& \displaystyle
	\sum_{m=0}^\infty \>
	    (a_m + \gamma b_0 \, a_{m{-}1}) \, b_0^{-m} \,
	    (t/ \Lambda^2)^n \,
	    n^{m+n\gamma-1} \, \Gamma (1{-}m{-}n\gamma, ny)
\nonumber
\\
    &\sim& \displaystyle
	\sum_{m=0}^\infty \>
	    (a_m + \gamma b_0 \, a_{m{-}1}) \, b_0^{-m} \,
	    \sum_{k=0}^\infty \,
		(-)^k
		{\Gamma (m{+}n\gamma{+}k) \over \Gamma (m{+}n\gamma)} \,
		n^{-k-1} y^{-m-k} e^{i n \phi} \,,
\end {eqnarray}
where $a_{-1} \equiv 0$, and  the asymptotic expansion of the
incomplete gamma function has been used,
\begin{equation}
    \Gamma (1{-}\alpha,x)
    \equiv
	\int_x^\infty dz \, z^{-\alpha} e^{-z}
    \sim
	x^{-\alpha} e^{-x}
	\sum_{k=0}^\infty \, {(-)^k \over x^k}
	{\Gamma(\alpha{+}k) \over \Gamma (\alpha)} \,,
\end{equation}
which is valid for $|x| \to \infty$ with $\arg(x) < 3\pi/2$.

    When evaluating the low momentum contribution,
the asymptotic expansion of the spectral density
can only be used when $s \gg \Lambda^2$.
However, for $n > 0$, the integrand of $I_n^-(t)$
is strongly peaked about the upper limit $s = |t|$.
Hence, one may ignore the contribution to the integral from
low momenta, $s < \kappa$, for some cutoff $\kappa$ chosen to
scale with $t$ so that
$\kappa/|t|$ vanishes faster than any power of $g^2(t)$
while simultaneously $g^2(\kappa) \to 0$ as $|t| \to \infty$.
Thus, for $n > 0$,
\begin {eqnarray}
    I_n^-(t) &=&
	\int_\kappa^{|t|} {ds \over s} \, \rho (s) \, (s/t)^n + O(\kappa/t)^n
\nonumber
\\
    &\sim& \displaystyle
	\int_{w(\kappa)}^y dz \> (1 + \gamma / z) \,
	\left( \sum_{m=0}^\infty a_m \, (b_0 z)^{-m} \right)
	(t/ \Lambda^2)^{-n} \,
	z^{n \gamma} \, e^{n z}
\nonumber
\\
    &\sim& \displaystyle
	\sum_{m=0}^\infty \>
	    (a_m + \gamma b_0 \, a_{m{-}1}) \, (-b_0)^{-m} \,
	    (t/ \Lambda^2)^{-n} \,
	    e^{i\pi n \gamma} \,
	    n^{m-n\gamma-1}
\nonumber
\\
    && \quad\qquad \times \;
	    \left[ \Gamma (1{-}m{+}n\gamma, e^{-i\pi} ny) -
	     \Gamma (1{-}m{+}n\gamma, e^{-i\pi} nw) \right]
\nonumber
\\
    &\sim& \displaystyle
	\sum_{m=0}^\infty \>
	    (a_m + \gamma b_0 \, a_{m{-}1}) \, b_0^{-m} \,
	    \sum_{k=0}^\infty
		{\Gamma (m{-}n\gamma{+}k) \over \Gamma (m{-}n\gamma)} \,
		n^{-k-1} y^{-m-k} e^{-i n \phi} \,,
\end {eqnarray}
where $w(\kappa) \equiv 1 / (b_0 \, g^2(\kappa))$.

    To evaluate the final term $I^-_0(t)$, define a ``subtracted'' spectral
density which is integrable from $0$ to $\infty$,
\begin {equation}
    \bar \rho(s) \equiv \rho(s) - \theta (s{-}\mu^2) \, (a_0 + a_1
	g^2(s)) \,,
\end {equation}
where $\theta(x)$ is the unit step function, and let
$
    C(\mu^2) = \int_0^\infty (ds / s) \, \bar \rho (s)
$.
Then,
\begin {eqnarray}
    I^-_0(t) &\equiv&
	C(\mu^2)
	+ \int_{\mu^2}^{|t|} {ds \over s} \, (a_0 + a_1 g^2(s))
	- \int_{|t|}^\infty {ds \over s} \, \bar \rho (s)
\nonumber
\\
    &\sim&
	C(\mu^2)
	+ a_0 \ln (|t|/\mu^2)
	+ a_1 \int_{w(\mu^2)}^y dz \> (1 + \gamma / z) \, (b_0 z)^{-1}
\nonumber
\\
    && \qquad {}
	-\sum_{m=2}^\infty \> a_m
	\int_y^\infty dz \> (1 + \gamma / z) \,
	(b_0 z)^{-m}
\nonumber
\\
    &\sim&
	C + a_0 y
	+ (a_1 + \gamma b_0 \, a_0) \, b_0^{-1} \, \ln y
\nonumber
\\
    && \qquad {}
	-\sum_{m=1}^\infty \> (a_{m{+}1} + \gamma b_0 \, a_m) \,
	    b_0^{-m-1} \, y^{-m} / m \,,
\end {eqnarray}
where
$
    C \equiv
	C(\mu^2) - a_0 \ln (\mu^2 / \bar\Lambda^2)
	+ a_1 b_0^{-1} \,
	\bigl(\ln (b_0 g^2(\mu^2)) + \gamma b_0 g^2 (\mu^2)\bigr)\,
$
is (despite appearances) a $\mu$-independent constant.

    Finally, putting these results together, we find that
\begin {equation}
    \Delta K (t e^{i \phi}) \sim
    \tilde c_{-1} / g^2(t) + \tilde c_0 \, \ln (g^2(t))
    + \sum_{n=0}^\infty \> c_n(\phi) \, g^2(t)^n \,,
\end {equation}
where
\begin {eqnarray}
    \tilde c_{-1} &=& (\pi b_0)^{-1} \, (-a_0) \,,
\\
    \tilde c_0    &=& (\pi b_0)^{-1} (a_1 - \lambda \, a_0) \,,
\\
    c_0(\phi)
  &=& (\pi b_0)^{-1} (a_1 {-} \lambda \, a_0)	\, \ln (b_0)
		- {1\over\pi}(a_0 \, \Delta_{0,0} (-\phi) + C)  \,,
\\
    c_n(\phi)
  &=& (\pi b_0)^{-1} {1\over n}(a_{n+1} {-} \lambda \, a_n)
		       - {1 \over \pi} \sum_{k=0}^n
			    (a_{n-k} {-} \lambda \, a_{n-k-1}) \, b_0^k \,
			    \Delta_{k,n-k} (-\phi) \,,
			    \label {two-term-c_n}
\end {eqnarray}
and
\begin {equation}
    \Delta_{k,m} (\phi) \equiv \sum_{{n=-\infty \atop n \ne 0}}^\infty \>
		e^{i n \phi} \, n^{-k-1} \,
		{ \Gamma (m+k+n\lambda/b_0) \over \Gamma (m+n\lambda/b_0)} \,.
\end {equation}
For $k \ge 0$, $\Delta_{k,m}(\phi)$ is a polynomial in $\phi$ of order $k+1$.
Inserting the expansion (\ref {I^n_ml}) of the ratio of gamma functions,
noting that
\begin {equation}
\sum_{{n=-\infty \atop n \ne 0}}^\infty \>
		e^{i n \phi} \, n^{-k-1} \,
    =
    - { (2\pi i)^{k+1} \over (k{+}1)!} \,
    B_{k+1}(\phi/2\pi) \,,
\end {equation}
where $B_k(x)$ is a Bernoulli polynomial,
and evaluating the result at $\phi = \pi$
yields the same result (\ref {c_n = sum a_n}) found earlier.

    The inverse to these relations, which will express the absorptive
coefficients in terms of the dispersive coefficients,
may be derived in a brute-force fashion by integrating the recursion relations
\begin {equation}
    i {d \over d\phi} c_{n+1}(\phi) =
    - b_{n+1} \, \tilde c_{-1} + b_n \, \tilde c_0
    + \sum_{m=1}^n m \, b_{n-m} \, c_m (\phi)
\label {recursion-rel}
\end {equation}
previously derived in Sec.~III [Eq.~(\ref {theta-dep}) with $\theta =
\phi - \pi$].
Since there is no explicit $\phi$-dependence in these relations,
one easily sees that $c_n(\phi)$ must be an $n$-th order polynomial
in $\phi$.
If the initial conditions are known values of the Euclidean coefficients,
$\{ c_n \equiv c_n (\phi=\pi) \}$, then a general form for
the $\phi$-dependence
which reproduces the initial conditions is
\begin {equation}
    c_n (\pi{-}\phi) = c_n + \sum_{l=1}^n \>
			    K_{n,l} \,
			    {(i b_0 \phi)^l \over l!} \,
\end {equation}
Substituting this form into the recursion relations (\ref {recursion-rel})
completely determines the coefficients $K_{n,l}$.
For our beta function with $ b_n = b_0 \, \lambda^n$, one finds that
\begin {equation}
    K_{n+1,l+1} =
	(\tilde c_0 - \lambda \tilde c_{-1}) \,
	\lambda^{n-l} \, I^n_{0,l}
    + \sum_{m=1}^{n-l}
	\, m \, c_m \, \lambda^{n-l-m} \, I^n_{m,l} \,,
\end {equation}
where $\{ I^n_{m,l} \}$ denotes the same combinatoric factors
\begin {equation}
    I^n_{m,l} \equiv
    \sum_{k_1 = m+1}^{n-l+1}
    \sum_{k_2 = k_1+1}^{n-l+2}
    \cdots
    \sum_{k_l = k_{l-1}+1}^{n}
    k_1 k_2 \cdots k_l
\end {equation}
whose generating function was introduced in (\ref {I^n_ml}).
Evaluating this result at $\phi = \pi$ and taking the imaginary part
immediately yields the expression for the absorptive coefficients
quoted in Eq.~(\ref {a_n = sum c_n}).

\appendix{known results}       \label{known}

Here we collect previous perturbative results and present them in terms of
our notation.
In the modified-minimal subtraction ($\overline{{\rm
MS}}$) scheme which uses a coupling that we now denote by $\bar g^2$,
the known QCD corrections to the $R$-ratio have the form
\begin{equation}
    R(s) = 3 \biggl( \sum \nolimits_f Q_f^2 \biggr)
		\left [1 + {{\bar g}^2(s) \over 4 \pi^2}
		+ r_2 \left ({\bar g}^2(s) \over 4 \pi^2 \right )^2
		+ r_3 \left ({\bar g}^2(s) \over 4 \pi^2 \right )^3
		+ O({\bar g}^8) \, \right],
\label{ASY}
\end{equation}
where $Q_f$ is the charge of the quark of flavor $f$.
The first two terms are not difficult to calculate
(see for example \cite{Brown}).
The coefficient $r_2$ for the third term has been obtained independently
by several groups \cite{Chetyrkin,Dine,Celmaster},
\begin{equation}
    r_2 = 1.9857 - 0.1153 \, N_f \,,
\end{equation}
where $N_f$ is the total number of quark flavors.
Recently, two groups \cite{Larin,Samuel}
have calculated the fourth coefficient,
\begin{equation}
    r_3 = -6.6368 - 1.2001 \, N_f - 0.0052 \, N_f^2 - 1.2395
	  \, {(\sum_f Q_f)^2 \over (3 \sum_f Q_f^2)} \,.
\label{result}
\end{equation}
The beta function ${\bar \beta}({\bar g}^2) \equiv \mu^2 d \bar g^2 / d \mu^2$
in the $\overline{{\rm MS}}$-scheme
has the following asymptotic expansion
\begin{equation}
    {\bar \beta} ({\bar g}^2) = - b_0 {\bar g}^4
	- b_1 {\bar g}^6 - b_2 {\bar g}^8 - O({\bar g}^{10}) \,,
\label{msbeta}
\end{equation}
where
\begin {eqnarray}
    4 \pi^2 b_0 &=& {1 \over 4} \left(11 - {2 \over 3}N_f \right) ,
\\
\noalign {\hbox {while \cite{Ca,Jo}}}
    (4 \pi^2)^2 b_1 &=& {1 \over 16}
		\left(102 - {38 \over 3} N_f \right)  ,
\\
\noalign {\hbox {and \cite{Tarasov}}}
    (4 \pi^2)^3 b_2 &=& {1 \over 64} \left({2857 \over 2} -
		{5033 \over 18} N_f + {325 \over 54} N_f^2 \right) .
\end {eqnarray}

To apply our results, we need first to express the $\overline{{\rm
MS}}$ expansion in terms of our coupling  $g^2$ which has
a two-term inverse beta function.
The two couplings are related by
\begin{equation}
    {\bar g}^2 = g^2 + \alpha_2 g^4 + \alpha_3 g^6 + O ({g}^8) \,,
\label{newcpl}
\end{equation}
and
\begin{equation}
    {\bar \beta} ({\bar g}^2(g^2)) =
	\beta (g^2) {d {\bar g}^2(g^2) \over d g^2} \,.
\label{bbrel}
\end{equation}
Requiring that
\begin{equation}
    \mu^2 {d g^2 \over d \mu^2}
    \equiv
    \beta (g^2)
     = - {b_0 g^4 \over 1 - \lambda g^2} \,,
\end{equation}
and inserting the expansions (\ref{msbeta}) and (\ref{newcpl})
into the relation (\ref{bbrel}) yields
\begin{equation}
    \alpha_3 = {b_2 \over b_0} - \lambda^2 + \alpha_2( \alpha_2 +
\lambda ) \,,
\end{equation}
with no constraint on the first parameter $\alpha_2$.
Since we have not yet specified the renormalization point for
the coupling $g^2$,
there is a one-parameter family of couplings
with a two-term inverse beta function.

We shall require that $\alpha_2$ vanishes
so as to keep $g^2$ as close to ${\bar g}^2$ as possible.
This redefinition then shifts only the fourth term in the expansion of
$R(s)$,
\begin{equation}
    R(s) = 3 \biggl( \sum \nolimits_f Q_f^2 \biggr)
	\left [
	    1 + {g^2 \over 4 \pi^2}
	    +  r_2 \left ({g^2 \over 4 \pi^2} \right )^2
	    + (r_3 + 16 \pi^4 \alpha_3) \left({g^2 \over 4 \pi^2} \right)^3
	    + O(g^8) \,
	\right ] \!.
\label{scaled}
\end{equation}
Since $R(s) = 12 \pi \, {\rm Im} K (s + i0^+)$, the first four
absorptive coefficients are
\begin{eqnarray}
    a_0  &=& {1 \over 4 \pi} \sum \nolimits_f Q_f^2 \,,
\nonumber\\
    4 \pi^2\, a_1 &=& a_0 \,,
\nonumber\\
    (4 \pi^2)^2 \, a_2 &=& r_2\,a_0 \,,
\nonumber\\
    (4 \pi^2)^3 \, a_3 &=& (r_3 + 16 \pi^4 \alpha_3) a_0 \,.
\end{eqnarray}
Relations (\ref{done}) and (\ref{finished}) determine the first four
dispersive coefficients in terms of the first four absorptive
coefficients,
\begin{eqnarray}
    \tilde c_{-1} &=& - {a_0 \over \pi b_0} \,,
\nonumber\\
    \tilde c_0 &=& {a_1 - \lambda a_0 \over \pi b_0} \,,
\nonumber\\
    c_1 &=& {a_2 - \lambda a_1 \over \pi b_0} \,,
\nonumber\\
    c_2 &=& {a_3 - \lambda a_2 \over 2 \pi b_0}
	 + {\pi b_0 a_1 \over 6} \,.
\end{eqnarray}
For five flavors, $N_f= 5$, the numerical values of the parameters are
\begin{eqnarray}
    4 \pi^2 b_0 &=& 1.917 \,,\qquad
    \lambda = b_1 / b_0 = 0.03194 \,,
\nonumber\\
    r_2 &=& 1.409 \,, \qquad r_3 = - 12.81 \,,
\nonumber\\
    (4 \pi^2)^2 \alpha_3 &=& - 0.12 \,.
\end{eqnarray}
The values of the first four absorptive coefficients become
\begin{eqnarray}
    a_0 &=& 0.09726 \,,
\nonumber\\
    4 \pi^2 a_1 &=&  a_0 \,,
\nonumber\\
    (4 \pi^2)^2 a_2 &=& \phantom- 1.409 \, a_0 \,,
\nonumber\\
    (4 \pi^2)^3 a_3 &=& - 12.92 \, a_0 \,,
\end{eqnarray}
and the first four dispersive coefficients are given by
\begin{eqnarray}
    \tilde c_{-1} &=& -6.556 \, a_0 \,,
\nonumber\\
    \tilde c_0 &=& -0.04333 \, a_0 \,,
\nonumber\\
    4 \pi^2 c_1 &=& \phantom- 0.02463 \, a_0 \,,
\nonumber\\
    (4 \pi^2)^2 c_2 &=& -0.2168 \, a_0 \,.
\end{eqnarray}
The presence of the term $(4 \pi^2)^2 \, \alpha_3$, caused by
the difference between our coupling  $g^2$ and
the more conventional $\overline{{\rm MS}}$ coupling  ${\bar g}^2$,
produces only a small (1\%) change in the fourth expansion
coefficient of the $R$-ratio (\ref{scaled}).

    Given these results, one may compute the first three partial sums
in the modified Borel transform of the absorptive coefficients,
\begin {equation}
    {\cal A}_K (z) \equiv \sum_{n=1}^K \>
    {\Gamma (1 {+} \lambda z) \over \Gamma (n {+} 1 {+} \lambda z)} \>
    n \, a_n \, z^n \,.
\end {equation}
The results are plotted in Fig.~1 for two different numbers of quark flavors.
The series of partial sums is obviously highly unstable when $b_0 z$ is
less than about $-1/2$.
Based on these graphical results, the presence of an ultraviolet
renormalon singularity at $b_0 z = -1$ is certainly not surprising.
Most intriguing, however, is the behavior of these partial sums
near $b_0 z = +1$.
As discussed in Sec.~II,
unless the absorptive transform ${\cal A}(z)$ vanishes at $b_0 z = 1$
the dispersive Borel transform will be singular at $b_0 z = 1$,
leading to previously unknown non-perturbative $1/q^2$ corrections
in the operator product expansion.
If the complete ${\cal A}(z)$ does have a zero at $b_0 z = 1$,
then in the sequence of partial sums one might hope to see a zero
on the positive axis whose position converges to $b_0 z = 1$.
As Fig.~1 shows, the differences between the partial sums grow
as $z$ increases from zero.
However, for one quark flavor, at $b_0 z = 1$ the first two partial
sums differ from the third by only $-21$\% and $+19$\%, respectively.
While hardly conclusive, the data for one flavor appear to support
the simplest hypothesis:
that further partial sums will converge to a non-zero value at $b_0 z = 1$,
leading one to predict the existence of non-perturbative $1/q^2$
effects.
As the number of quark flavors increase, the partial sums (for fixed
values of $b_0 z$) become increasingly unstable.
For $N_f = 5$, the last partial sum does have a zero near $b_0 z = 1.3$,
but the differences between the different partial sums are clearly too large
to draw any meaningful conclusion about the true value at $b_0 z = 1$.

\begin {figure}
    {
    
    \leavevmode \hspace {0.6in}
    \setlength {\unitlength}{1cm}
    \begin {picture}(0,9)(0,0)
    \thicklines
    \put (11.45,3.99){\vector (1,0){0.3}}
    \put (12.0,3.99){\makebox(0,0)[l]{$b_0 z$}}
    \put (4.27,7.75){\vector (0,1){0.3}}
    \put (4.27,8.4){\makebox(0,0)[b]%
	{$\quad 4 \pi^2 \, {\cal A}_K(z) \, / \sum_f Q_f^2$}}
    \put (10,8.4){\makebox(0,0)[l]{(a) $N_f = 1$}}
    \end {picture}
    \epsfbox {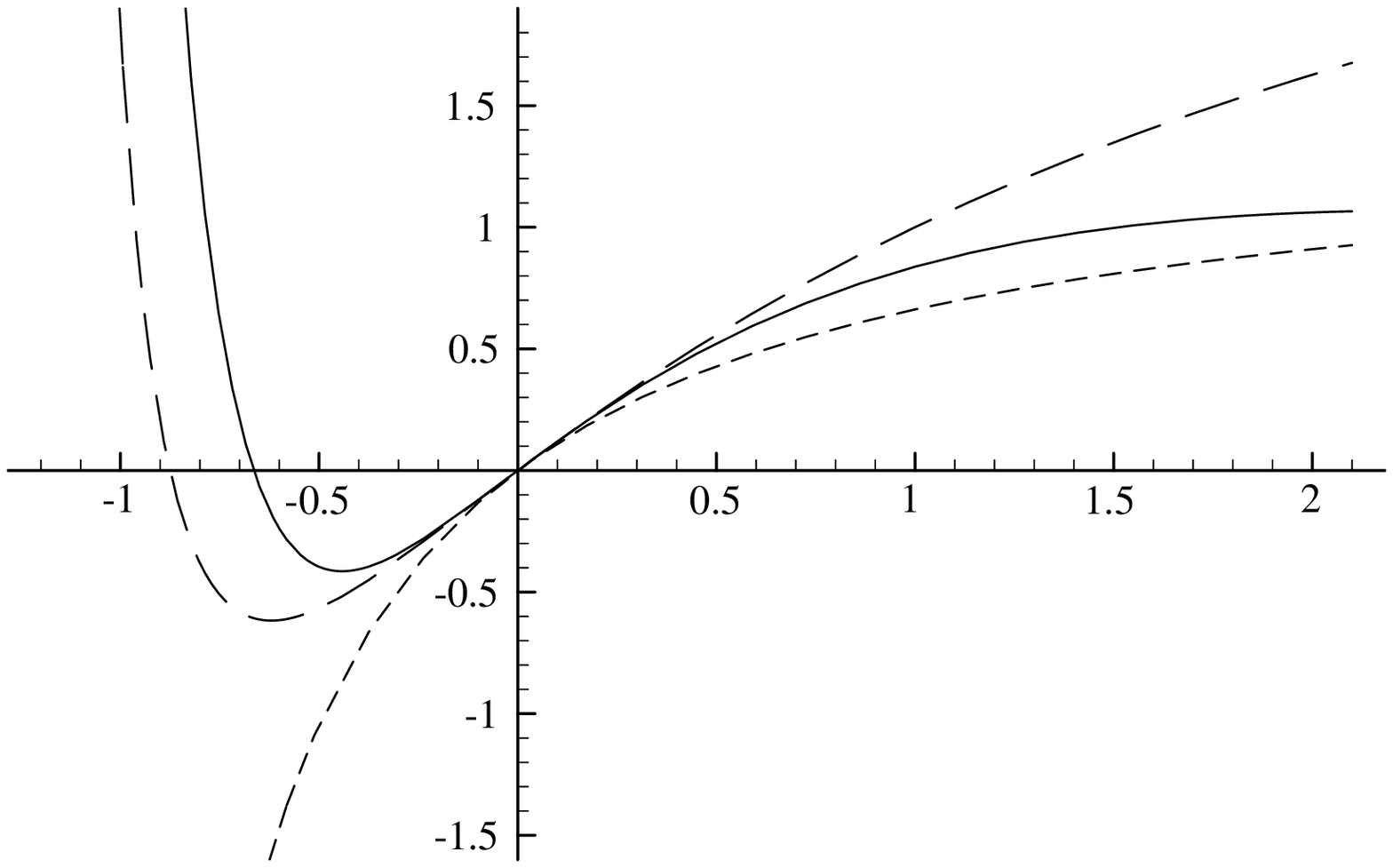}
    \vspace* {0.5cm}

    \leavevmode \hspace {0.6in}
    \begin {picture}(0,9)(0,0)
    \thicklines
    \put (11.45,3.99){\vector (1,0){0.3}}
    \put (12.0,3.99){\makebox(0,0)[l]{$b_0 z$}}
    \put (4.27,7.75){\vector (0,1){0.3}}
    \put (4.27,8.4){\makebox(0,0)[b]%
	{$\quad 4 \pi^2 \, {\cal A}_K(z) \, / \sum_f Q_f^2$}}
    \put (10,8.4){\makebox(0,0)[l]{(b) $N_f = 5$}}
    \end {picture}
    \epsfbox {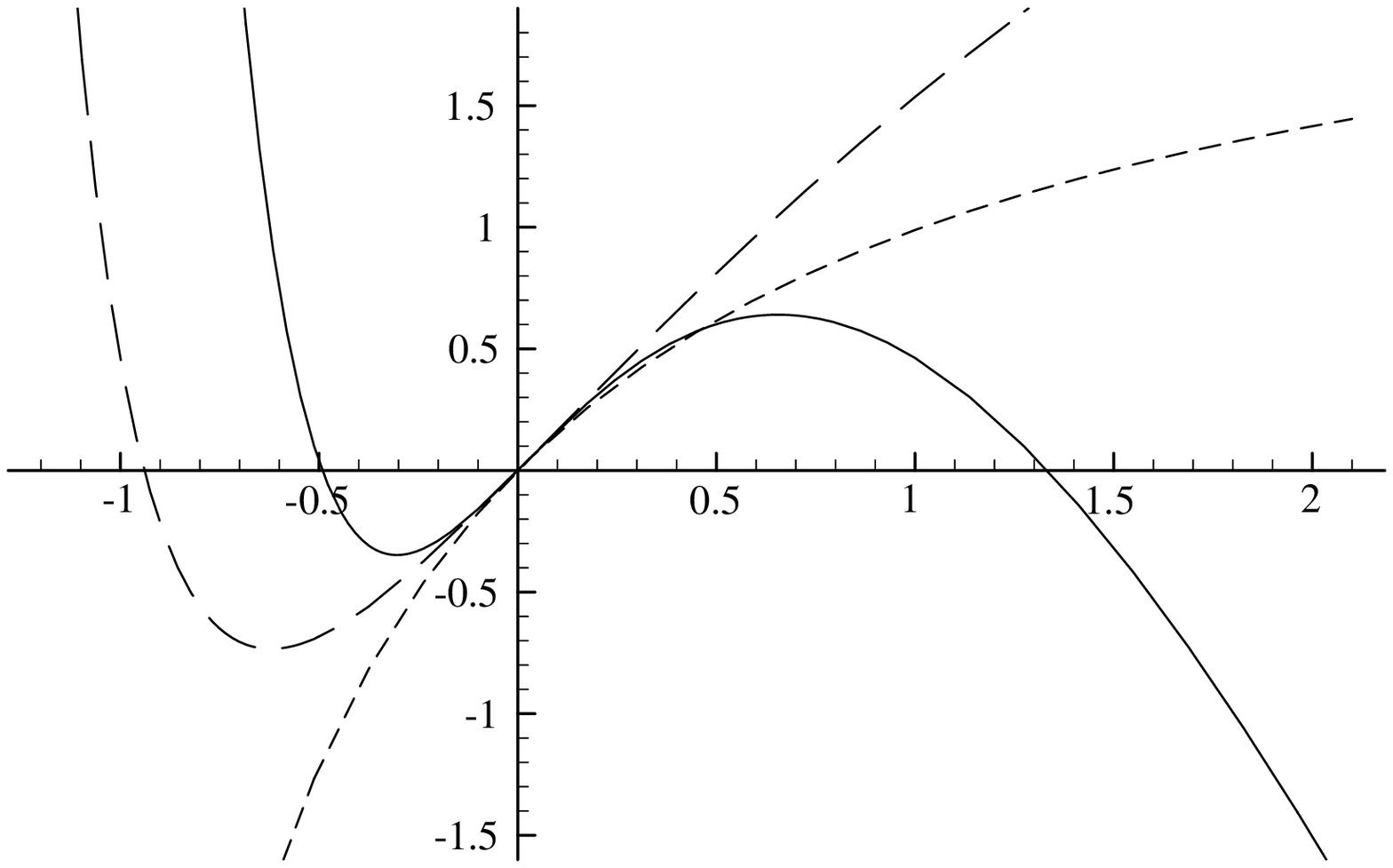}
    \caption
	{%
	\advance\baselineskip by -8pt
	First three partial sums for the absorptive modified Borel transform
	${\cal A}(z)$ for $N_f = 1$ and $N_f = 5$.
	Plotted is $4 \pi^2 \, {\cal A}_K(z)/ \sum_f Q_f^2$ versus $b_0 z$.
	The short dash, long dash, and solid lines denote the first, second,
	and third partial sums, respectively.
	}
    }
\end {figure}

\end {document}